\documentstyle[epsfig,preprint,aps]{revtex}
\begin{document}
\tighten
\preprint{
\vbox{
\hbox{ADP-01-35/T467}
}}
\draft
\title{Structure Functions of Unstable Lithium Isotopes}
%
\author{K. Saito$^a$\footnote{E-mail address: ksaito@nucl.phys.tohoku.ac.jp}, 
M. Ueda$^b$\footnote{mueda@nucl.ph.tsukuba.ac.jp}, 
K. Tsushima$^c$\footnote{ktsushim@physics.adelaide.edu.au. Address from 
November 1, 2001; Department of Physics and Astronomy, 
University of Georgia, Athens, GA 30602, USA 
(tsushima@hal.physast.uga.edu)} 
and A.W. Thomas$^c$\footnote{athomas@physics.adelaide.edu.au}}
\address{$^a$ Tohoku College of Pharmacy, Sendai 981-8558, Japan \\ 
$^b$ Institute of Physics, University of Tsukuba, Tsukuba 305-8571, Japan \\ 
$^c$ Department of Physics and Mathematical Physics and \\
Special Research Center for the Subatomic Structure of Matter (CSSM) \\
Adelaide University, Adelaide SA 5005, Australia}
\maketitle
\begin{abstract}
We study both the spin-average and spin-dependent structure functions of 
the lithium isotopes, $^{6-11}$Li, which could be measured at 
RIKEN and other nuclear facilities in the future.  
It is found that the light-cone momentum 
distribution of the valence neutron in the halo of $^{11}$Li is very 
sharp and symmetric around $y = 1$, because of the weak binding. This implies 
that such neutrons are in an environment very close to that of a free 
neutron.  The EMC ratios for Li isotopes are then calculated. 
Furthermore we investigate a new ratio, $R_d$, of the difference between 
the Li structure functions of mass number $A$ ($F_2^A$) and
$A-1$ ($F_2^{A-1}$) to the difference between the structure 
functions of the deuteron 
and free proton. We study the 
possibility of extracting the neutron structure function 
from data for the nuclear structure functions of the Li isotopes. 
In particular, the ratios $R_n$ for $ A = 9$ and $11$ present an   
attractive possibility for extracting the free neutron structure function.  
Next we calculate 
the spin-dependent structure functions of $^{7,9,11}$Li isotopes, which 
have spin of 3/2. For such nuclei the spin structure function is given 
in terms of the multipole spin structure functions; $^{3/2}_{~1}g_1$, which 
is analogous to 
the usual spin structure function, $g_1$, for a target with spin 1/2,
and a new one, $^{3/2}_{~3}g_1$, which first arises for a target with
spin 3/2. 
The effect of the nuclear binding and Fermi motion on $^{3/2}_{~1}g_1$ is 
about $10 \%$ in the region $x < 0.7$, but 
it becomes quite important at large $x$. 
The spin structure function of $^{3/2}_{~3}g_1$ is negative at small $x$ 
but it becomes positive in the region $0.2 < x < 0.5$. However, the 
magnitude is very small. 
At large $x$ it is again negative and its absolute value becomes large 
because of the Fermi motion.  Finally, we discuss the modification 
of the Gottfried and Bjorken integrals in a nuclear medium and point 
out several candidates for a pair of mirror nuclei to study the 
{\em flavor-nonsinglet} quark distributions in nuclei. 
\end{abstract}

\newpage
\section{Introduction}
\label{sec:intro}

The recent development of secondary radioactive beams with sufficient
intensities, which are produced through high energy fragmentation of nuclei, 
has opened a new frontier in nuclear physics, i.e., the study of  
nuclei with extreme ratios of the numbers of neutrons, $N$, to the number of 
protons, $Z$. Such nuclei, far from the stability line,
have been investigated
intensively at many nuclear facilities in the world
as one of the main subjects of nuclear science today~\cite{tani1}. 
There are various of interesting problems such as 
weakly bound nuclei, low-density nuclear matter, 
novel nuclear structures of halo and skin,
very asymmetric nuclear matter, appearance/disappearance of magic numbers
in the vicinity of the neutron and proton drip line~\cite{tani1,muses}. 

In particular, it is noticeable that the halo structure in some light nuclei
was discovered on and close to the neutron drip line~\cite{tani2}.
Those nuclei have spatial extensions much larger than naive expectations  
based on their mass numbers.  
A typical example is $^{11}$Li, which has a root mean 
square (rms) radius of about 3.1 fm. 
In Table~\ref{tab:exp} we summarize the observed data of 
$^{6-11}$Li isotopes~\cite{data,69li}. 
The experimental data suggests that nuclei like $^{11}$Li 
could be understood as a core surrounded by one or two more 
neutrons, which extend to several times the nuclear radius -- the so-called 
{\em halo} nuclei. This picture is also supported by recent experiments 
with break-up of loosely bound neutron-rich projectiles on light 
targets~\cite{tani3}. 

The $\beta$-unstable nuclei in the vicinity of 
the neutron drip line have two distinctive features.
The first feature is a large asymmetry in $Z$ and $N$. 
For stable nuclei with $Z \geq 3$, due to the Coulomb 
interaction and the nuclear saturation property,
(i) the ratio $N/Z$ is 1.0 $\sim$ 1.6, 
(ii) the observed central density is $\sim 0.17$ fm$^{-3}$
except for very light nuclei, and (iii) the density distributions of 
protons and neutrons are similar and the thickness of the nuclear
surface is almost unchanged for all nuclei. For $\beta$-unstable nuclei 
close to the neutron drip line, the $N/Z$ ratio is about 2 $\sim$ 
4 and the distributions of protons and neutrons are considerably decoupled
(i.e., neutron halo and neutron skin). 

The second feature is a small separation energy of 
the valence neutron. For light stable nuclei, where $N \approx Z$,
the separation energy of the last nucleon is usually several MeV. On the 
other hand, with an increasing number of neutrons the separation energy 
of the last neutron, $S_n$, varies from several MeV 
to less than 1 MeV at the neutron drip line, while the separation energy
of a proton tends to increase and sometimes exceeds 10 MeV
near the neutron drip line. This difference between the separation energy 
of a neutron and that of a proton implies a difference in the Fermi energies 
of neutrons and protons, and it provides an understanding of the mechanism of
formation of neutron halos and skins as well as the instability
of $\beta$-unstable nuclei. 

The partonic distribution functions in such nuclei are of considerable 
interest in view of these features~\cite{sai1}.  
The distributions of quarks in a nucleus differ significantly from 
those in the free nucleon -- the so-called European Muon Collaboration 
(EMC) effect~\cite{review,pw}. Since the early 80's a large amount of data on 
{\em flavor-singlet} nuclear structure functions, i.e., $N \simeq Z$, 
has been collected. However, relatively less attention has been paid to the 
{\em flavor-nonsinglet} structure functions of a nucleus.  Measurements of 
nonsinglet nuclear structure functions in deep inelastic 
scattering (DIS) could shed light on 
many phenomena involving nonperturbative QCD in nuclei, such as SU(3) 
symmetry breaking, flavor asymmetry~\cite{kumano} and so on. 

Recently we have studied the nonsinglet structure function of the 
lightest mirror nuclei ($^3$He and $^3$H)~\cite{sai1,vadim} and pointed out 
that the nonsinglet structure function in those nuclei is enhanced at 
small $x$ by nuclear shadowing. Furthermore, we suggested that the 
Gottfried integral is generally divergent because of charge 
symmetry breaking. We have also investigated the effect of the medium 
modification of pion fields on the {\em flavor-nonsinglet} structure function 
of the lightest mirror nuclei~\cite{sai2}. It is intriguing 
to study such nonsinglet structure functions using a pair of {\em unstable} 
nuclei~\cite{sai1} because such nuclei could give rise to a large 
asymmetry in isospin space -- recall the first feature for unstable nuclei 
discussed above.  The current rapid development of 
radioactive nuclear beam techniques at many nuclear facilities 
in the world will provide good opportunities for advancement to
the study of nonsinglet structure functions 
in the near future~\cite{muses,future}. 

Utilizing the second feature of neutron-rich nuclei far from the
stability line, i.e., a wide range of the 
neutron separation energy, we could investigate the binding  
(off-mass shell~\cite{off} and shell structure dependences) and Fermi motion 
effects on nuclear structure functions~\cite{sai1}. Furthermore, the 
neutron structure function, $F_2^n$, could be extracted from nuclear 
structure functions of neutron-rich nuclei, by virtue of the weakness of 
the neutron separation energy~\cite{sai1}. 

It is very important to determine the neutron structure function 
from nuclear experimental data. 
The proton structure function, $F_2^p$, which at intermediate 
and large-$x$ measures a charge-squared
weighted combination of the valence $u$ and $d$ distributions, depends 
mostly on the $u$ quark distributions 
owing to the larger weight of the $u$ quark charge. It is therefore necessary 
to know the neutron structure function as precisely as
possible to determine the individual isospin distributions separately. 

The problem of extracting neutron structure functions from nuclear data 
is rather old~\cite{d1}. 
Because of the absence of a free neutron target the deuteron is 
traditionally used as an effective neutron target: $F_2^n \approx 
F_2^d - F_2^p$ ($F_2^d$ being the deuteron structure function).  
While this approximation may be valid for $0.1 < x < 0.4$, it breaks 
down dramatically for large $x$ because of the Fermi motion. 
The Adelaide group~\cite{mel} has recently 
reanalyzed the deuteron structure function to determine $F_2^n$, and 
they showed that the nuclear binding and Fermi motion effects 
were significantly larger than those in earlier analyses. In particular, 
ignoring the effect of nuclear binding yields an error of 
up to 50\% in the ratio $F_2^n/F_2^p$ extracted at $x \sim 0.75$. 

Recently, the $A=3$ system ($^3$He and $^3$H) is also being seriously 
considered as an alternative way to extract the neutron structure 
function~\cite{bissey,pace}. Using realistic Faddeev wave functions, 
it was shown that $F_2^n$ could be extracted from the ratio of those 
structure functions, 
where the nuclear effects are cancel each other so that 
the ratio is within 2\% of unity for $x < 0.85$. 
In the present paper we discuss another way to extract
the neutron structure function using the radioactive Li isotopes. This is 
based on the novel idea that the valence neutrons in the
$\beta$-unstable nuclei are so loosely bound that they may be regarded as
essentially as free neutrons. 

For the last decade polarized DIS experiments have yielded a number of 
important results concerning the spin structure of the nucleon. 
The measurements of the proton spin structure 
function, $g_1^p$, by EMC~\cite{spinEMC}, combined with flavor-nonsinglet 
matrix elements from weak decays, gave significant information on the 
singlet axial charge of the proton. The small size of this resulted in 
the so-called 
{\em proton spin crisis}, and it has been reanalyzed by many 
groups~\cite{recentspin}. As an extra source of information, 
it is important to measure the neutron spin structure function, $g_1^n$, 
for testing the fundamental Bjorken sum rule of the free nucleon. 
However, as in the unpolarized case, the absence of the free neutron target 
means that nuclei have to be used for this purpose.
In fact, to study $g_1^n$ the SLAC E142 Collaboration~\cite{SLACE142} 
has measured the spin structure function of $^3$He, while the Spin Muon
Collaboration (SMC)~\cite{SMC} has observed the $g_1$ structure
function of the deuteron. (For recent experiments, see Ref.~\cite{g1exp}.) 

The nuclear binding, Fermi motion and shadowing 
modify the parton distributions in a nuclear medium. 
As in the case of Gottfried sum rule in nuclei~\cite{sai1,vadim}, 
it is worthwhile to study how the {\em flavor-nonsinglet} 
combination of the spin structure functions, which appears 
in the Bjorken sum rule, is modified in a nucleus.  
We can discuss it again using a pair of polarized mirror nuclei 
-- the promising pairs may be ($^3$He, $^3$H)~\cite{spin3} and 
($^7$Li, $^7$Be)~\cite{Guzey} (we will discuss several candidates 
in Sec.~\ref{sec:summary}). If the spin structure functions of unstable 
nuclei were measured with a certain accuracy, one could obtain 
significant information on the spin structure of partons in nuclei. 

In general, the spin-average DIS cross section, ${\bar \sigma}$, and 
polarization asymmetry, $\Delta \sigma$, of a target nucleus 
with mass $M_T$ and spin $J$ in the scaling limit is given 
by~\cite{jaffe}  
\begin{eqnarray}
\frac{d {\bar \sigma}^J}{dxdy} = C x [1 + (1-y)^2]
       \sum_{L={\rm even}} {^J_LF_1(x)} \rho_L^J , \label{cs1} \\
\frac{d \Delta \sigma^J}{dxdy} = C x [1 - (1-y)^2]
       \sum_{L={\rm odd}} {^J_Lg_1(x)} \rho_L^J , \label{cs2} 
\end{eqnarray}
with a kinematical factor $C = e^4M_TE/2\pi Q^4$, 
where $E$ is the incident lepton energy, $q^2 (= - Q^2)$ 
the four momentum transfer 
squared and $y$ the ratio of the energy transfer to $E$ in the target 
rest frame.  $^J_LF_1$ and $^J_Lg_1$ are, respectively, {\em multipole} 
spin-average and {\em multipole} spin-dependent structure functions,
which are defined in terms of 
the target helicity-dependent structure functions, $F_1^{JH}$ and 
$g_1^{JH}$ ($H$ stands for helicity), and a projection matrix, $M_L^J$: 
\begin{equation}
^J_LF_1 = {\rm tr}[M_L^J F_1^J] , \ \ \ ^J_Lg_1 = {\rm tr}[M_L^J g_1^J]. 
\label{multifg}
\end{equation}
Here $F_1^J$ and $g_1^J$ are $(2J+1) \times (2J+1)$ diagonal matrices with 
their elements $(F_1^J)_{HH} = F_1^{JH}$ and $(g_1^J)_{HH} = 
g_1^{JH}$, respectively, and $M_L^J$ is defined as
\begin{equation}
(M_L^J)_{HH'} = (-)^{J-L} (JH, J-H | L0) \delta_{HH'} , \label{M}
\end{equation}
where $(JH, J-H | L0)$ is 
the Clebsch-Gordan coefficient. In Eqs. (\ref{cs1}) and
(\ref{cs2}) $\rho_L^J$ is a multipole projection of a spin density matrix, 
$\rho^J$. The multipole structure functions are irreducible 
under rotation. (Note that in the scaling limit $^J_LF_2$ is related to 
$^J_LF_1$ through the Callan-Gross relation.) 

For a target with spin $J=1$ the DIS cross section is given in terms of 
three non-vanishing multipole structure 
functions, $^1_0F_1$, $^1_1g_1$ and $^1_2F_1$, in the scaling limit.  
This case corresponds, for example, 
to the DIS on $^6$Li (see Table~\ref{tab:exp}). 
The first one is then related to the usual spin-average structure functions, 
$F_1$, through 
\begin{equation}
^J_0F_1(x) = \frac{1}{\sqrt{2J+1}} \sum_H [q_\uparrow^{JH}(x) 
+ q_\downarrow^{JH}(x)] = 2 \sqrt{2J+1} F_1(x) , \label{relation}
\end{equation}
where $q_s^{JH}$ is the distribution function of quarks with spin $s$ in 
the target.  (Note that all flavor labels and charges are suppressed.) 
The usual $F_2$ structure function can be obtained using the 
Callan-Gross relation. 
The second one corresponds to the spin structure function for a 
spin-1/2 target, $g_1$, and they are related as $^1_1g_1 = \sqrt{2} g_1$. 
The last one is the {\em quadrupole} structure function for 
a spin-1 target, which is sometimes called $b_1$~\cite{b1}.  

The $^8$Li target has spin-2 (see Table~\ref{tab:exp}). 
In such a case there are five non-trivial 
structure functions in the scaling limit: 
$^2_0F_1$, $^2_1g_1$, $^2_2F_1$, $^2_3g_1$ and $^2_4F_1$. 
Then, $^2_0F_1$, $^2_1g_1$ and $^2_2F_1$ are analogous to $F_1$, $g_1$ 
and $b_1$, respectively, for $J=1$. The structure function
$^2_3g_1$ is analogous to the one that first appears 
at $J=3/2$, $^{3/2}_{~3}g_1$ (see below).  
The last one, $^2_4F_1$, first arises for $J=2$. 

The spin of Li isotopes with odd $A$ is $J=3/2$ (see Table~\ref{tab:exp}). 
In this case there are 
four non-vanishing multipole structure functions in the scaling limit: 
$^{3/2}_{~0}F_1$, $^{3/2}_{~2}F_1$, $^{3/2}_{~1}g_1$ and $^{3/2}_{~3}g_1$. 
The first one is again proportional to the usual spin-average 
structure functions, and the second one is analogous to the 
quadrupole structure function, $b_1$, for a spin-1 target.
The last two are sensitive to quark spin asymmetries. $^{3/2}_{~1}g_1$ is the
analog to $g_1$ for a spin-1/2 target:
\begin{equation}
^{3/2}_{~1}g_1(x) = \frac{1}{\sqrt{20}} [3 q_\uparrow^{3/2 \, 3/2}(x)
- 3 q_\downarrow^{3/2 \, 3/2}(x) + q_\uparrow^{3/2 \, 1/2}(x) - 
q_\downarrow^{3/2 \, 1/2}(x)] , \label{relg1}
\end{equation}
while $^{3/2}_{~3}g_1$ is a new asymmetry: 
\begin{equation}
^{3/2}_{~3}g_1(x) = \frac{1}{\sqrt{20}} [q_\uparrow^{3/2 \, 3/2}(x)
- q_\downarrow^{3/2 \, 3/2}(x) - 3 q_\uparrow^{3/2 \, 1/2}(x) + 
3 q_\downarrow^{3/2 \, 1/2}(x)] . \label{relg3}
\end{equation}
It is important to keep in mind that all nucleons contribute to the 
spin-average structure functions but that only unpaired valence nucleons 
mainly contribute to the other multipole structure functions. 

In the present paper we concentrate on the spin-average structure functions 
of $^{6-11}$Li and the spin-dependent structure functions of $^{7,9,11}$Li,
and study those structure functions in detail. 
Measurements of those unpolarized and polarized structure functions could be
performed by the MUSES project proposed at RIKEN~\cite{muses} and/or
other nuclear facilities in the future~\cite{future,future2}. 
The outline of the present paper is as follows. 
In Sec.~\ref{sec:Li}, we summarize the 
main features of the lithium isotopes measured in experiments, and 
calculate their wave functions using Hartree approximation. The spin-average 
structure functions of Li are presented in Sec.~\ref{sec:Listrfn}. 
Moreover, we discuss how the neutron structure function can
be extracted from Li nuclei. The present way
will provide a new alternative to study the
neutron structure function using radioactive ions rich in neutrons.
In Sec.~\ref{sec:spin} we calculate the spin structure functions of 
the $^{7,9,11}$Li isotopes.  The multipole spin structure functions of 
$^{3/2}_{~1}g_1$ and $^{3/2}_{~3}g_1$ for a spin-3/2 target are discussed in 
detail. The effect of the nuclear binding and Fermi motion on 
$^{3/2}_{~1}g_1$ is 
less than $10 \%$ for $x < 0.7$, but it becomes important at large $x$,  
as in the EMC effect.
The spin structure function of $^{3/2}_{~3}g_1$ is novel and it first 
arises for a spin-3/2 target.  It is negative for small $x$ but
it turns to be positive for $0.2 < x < 0.5$.  For large $x$ it is 
again negative and the magnitude becomes large because of the Fermi
motion.  We summarize our results and 
discuss several candidates for testing the Gottfried and Bjorken sum rules 
in mirror nuclei in the last section. 

\section{Structure of Li isotopes}
\label{sec:Li}

We show the properties of the $^{6-11}$Li isotopes in 
Table~\ref{tab:exp}.  It is quite remarkable that 
the two-neutron separation energy, $S_{2n}$, decreases dramatically 
as the atomic mass number, $A$ = $Z + N$, increases and that 
at the neutron drip line $S_{2n}$ of 
$^{11}$Li reaches about 300 keV.  The one-neutron separation energy, $S_n$, 
also diminishes with increasing $A$ 
(although it shows a dependence on the shell structure), 
while the one-proton separation energy, $S_p$, increases, mainly
due to the attractive $p$-$n$ interaction in nuclei. 
It is also notable that the rms radius, $r_A$, of 
$^{11}$Li is about 3.1 fm, which is significantly larger than 
$A^{1/3} \sim 2.2$ fm. Furthermore,
we list the energy levels of the first $1/2^-$ states 
of $^7$Li, $^7$Be, $^9$Li, and $^9$Be in 
Table~\ref{tab:exc}~\cite{data,69li}, 
as these are expected to provide information on the spin-orbit force 
for the nucleons in the $p$-shell. 

\subsection{Model for Li isotopes}
\label{subsec:model}

During the last 15 years the structure of light $\beta$-unstable 
nuclei has been studied intensively. In particular, $^{11}$Li, which 
is located at the neutron drip line, has attracted a lot of attention. 
There two valence neutrons are so weakly bound to $^9$Li that they form
a halo structure. From this point of view, some simple cluster
models such as the di-neutron cluster model~\cite{dineutron}, 
where the valence neutrons are correlated strongly enough to be 
regarded as one particle, were first proposed, and those simple models 
succeeded in reproducing the qualitative features of $^{11}$Li. 
With increasing observed data on $^{11}$Li, it was, however, 
realized that these simple cluster models could not explain the measured 
experimental data. 
Thus, some authors started calculating the properties of $^{11}$Li
using a three-body model~\cite{halo}, which was supported by the fact that
$^{11}$Li is a Borromean system ($^{11}$Li = $^9$Li $+ n + n$), 
where all the two-body sub-systems are unbound. 
At present the three-body model is one of the best ways 
to describe the observed properties of $^{11}$Li.  However, speaking 
strictly, the three-body model is only valid for structureless particles. 
In this sense a standard model to describe $^{11}$Li 
has not yet been established. 

Our aim in the present paper is to investigate the nuclear
structure functions of the Li isotopes and, in particular, to determine 
the momentum distribution of each nucleon in these 
nuclei. Thus, the three-body model is not suitable for our purpose. 
In addition to this, since we consider that the separation 
energy is more significant for calculating nuclear structure functions
than the other properties of a nucleus, in the present paper we 
use a simple shell model with Hartree approximation to calculate 
single-particle wave functions of Li isotopes. 
Note that it implies that the valence neutrons correlate 
rather weakly in the halo of $^{11}$Li. 
In subsection~\ref{subsec:comments}, 
we will discuss effects of residual two-body interactions. 

In Hartree approximation the total Hamiltonian of a $A$-nucleon system 
is given in terms of a sum of nucleon kinetic energy and 
an effective one-body potential, $U$: 
\begin{equation}
H = \sum_{i=1}^A \biggl [ \, -\frac{\hbar^2}{2M} {\vec \nabla_i}^2
+U(\xi_i) \, \biggr ] , 
\label{H}
\end{equation}
where $M$ (= 939 MeV) is the nucleon mass, and $\xi_i$ denotes position 
${\vec r}_i$, spin ${\vec \sigma}_i$, 
and isospin ${\vec \tau}_i$ of the $i$-th nucleon. 
The $A$-body wave function, $\Psi(\xi_1, \cdots, \xi_i, \cdots \xi_A)$, 
is then given by the product of single-particle wave functions, 
$\phi_{\alpha_i} (\xi_i)$ ($\alpha_i$ stands for all quantum numbers 
associated with the $i$-th nucleon).
In this model the single-particle separation energy is simply given 
by $-\varepsilon_i$. 

Determining the consistent isoscalar and isovector mean field potentials
is very important for discussing not only neutron-rich nuclei but also
neutron stars~\cite{star}. However, it would need several elaborate 
experimantal and theoretical works to fix both the potentials uniquely.
In the present paper, therefore, we suppose as a {\it simple} and 
{\it effective} potential as possible. It is spherical and consisits of 
the nuclear, $U_N$, and Coulomb, $U_C$, parts:
\begin{equation}
U(A,Z,{\tau_z};r) = U_N(A,Z,{\tau_z};r) + U_C(A,Z;r)
\frac{1-{\tau_z}}{2} ,
\label{U}
\end{equation}
where $\tau_z$ is the $z$-component of isospin ($\tau_z = +1$ for neutron 
($n$) and $-1$ for proton ($p$)). 
Here the nuclear part is given by~\cite{BM}  
\begin{eqnarray}
U_N(A,Z,{\tau_z};r) &=& \frac{-U_0}{1+\exp{\biggl ( \frac{r-R}{a} \biggr )}}
+ r_0^2\,\, (\vec{\ell} \cdot \vec{s})\,\, 
\frac{1}{r} \frac{d}{dr} \left[
\frac{-U_{LS}}{1+\exp{\biggl ( \frac{r-R}{a} \biggr )}} \right] , 
\label{bmpot}
\end{eqnarray} 
with $R = r_0 \cdot (A-1)^{1/3}$. 
In Eq. (\ref{bmpot}) the first and second terms are the central and 
spin-orbit forces, respectively, and we further suppose that
the potential depths, $U_0$ and $U_{LS}$, depend on $A, Z$, 
and ${\tau_z}$. The geometrical parameters, $r_0$ and $a$,
are taken to be common to both the protons and neutrons, 
in order to reduce the number of
parameters for the single-particle potential. The last assumption is
also based on the fact that the $^{6-9}$Li isotopes are considered to
have the normal density distribution (see Table 2 and Fig. 1 of 
Ref.~\cite{69li}).  (As will be shown 
in Fig.~\ref{fig:density}, however, the present potentials yield
decoupled proton- and neutron-density distributions in $^8$Li and $^9$Li
nuclei. We will discuss the reason for this discrepancy in 
subsection~\ref{subsec:comments}.)

Concerning the Coulomb part, we assume that it is formed by
a Gaussian charge distribution with the charge of $Z-1$ and a radius 
parameter $a_c$, 
\begin{equation}
\rho_c(r) = (Z-1)e \cdot (a_c \sqrt{\pi})^{-3} \, e^{-(r/a_c)^2} .
\label{charge}
\end{equation}
We then obtain 
\begin{equation}
U_C(r) = \frac{(Z-1) e^2}{r} \, {\rm erf}\biggl (\frac{r}{a_c} \biggr ) ,
\label{erf}
\end{equation}
where ${\rm erf}(x)$ is the error function \cite{abramowitz}
and $a_c$ is determined from the rms radius of the Gaussian 
distribution
\begin{equation}
\langle r^2 \rangle_{C} = \frac{3}{2}a_c^2 . 
\label{charge2}
\end{equation}
We here use the experimental rms radii of the matter distributions of
the Li nuclei except for $^{11}$Li.  For $^{11}$Li, we use the matter 
distribution of $^9$Li 
as the rms radius of the charge distribution.

\subsection{Shell configuration}
\label{subsec:shell}

Assuming that identical nucleons in a given shell orbit are coupled 
to a spin-singlet state and that no two-body 
residual forces exist, we can fix 
the configurations for the ground states in the Li isotopes, based on 
the simple shell model. For example, one can write the shell configuration 
for $^9$Li as
\begin{displaymath}
(\pi 1s_{1/2})^2(\pi 1p_{3/2})^1 (\nu 1s_{1/2})^2(\nu 1p_{3/2})^4 \ . 
\end{displaymath}
We also assume that the valence neutrons in $^{11}$Li occupy the 
$\nu 1p_{1/2}$ orbit although higher configurations such as $1d_{5/2}$ and 
$2s_{1/2}$ are thought to be involved in 
forming a realistic halo structure. 
In our model the single-particle excitation energies of a neutron and  
proton are given by 
\begin{equation}
\Delta \varepsilon(n) = \vert \, \varepsilon_n(1p_{1/2})-
\varepsilon_n(1p_{3/2}) \, \vert , \hspace{1cm}
\Delta \varepsilon(p) = \vert \, \varepsilon_p(1p_{1/2})-
\varepsilon_p(1p_{3/2}) \vert , 
\label{excen}
\end{equation}
respectively, and the rms radii of $^{6-9}$Li and $^{11}$Li nuclei 
are given by
\begin{eqnarray}
r_{A=6-9}^2 &=& \frac{1}{A}
\biggl ( \, 2 \langle r^2 \rangle_{\pi 1s_{1/2}} 
+ 2 \langle r^2 \rangle_{\nu 1s_{1/2}} 
+ \langle r^2 \rangle_{\pi 1p_{3/2}} 
+ (A-5) \langle r^2 \rangle_{\nu 1p_{3/2}} \, \biggr ) ,
\label{rms9} \\
r_{A=11}^2 &=& \frac{1}{A}
\biggl ( \, 2 \langle r^2 \rangle_{\pi 1s_{1/2}} 
+ 2 \langle r^2 \rangle_{\nu 1s_{1/2}} 
+ \langle r^2 \rangle_{\pi 1p_{3/2}} 
+ 4 \langle r^2 \rangle_{\nu 1p_{3/2}} 
+ 2 \langle r^2 \rangle_{\nu 1p_{1/2}} 
\, \biggr ) ,
\end{eqnarray}
respectively.

\subsection{Model parameters and comments}
\label{subsec:comments}

In order to obtain the potential parameters, we divide the 
Li isotopes into two groups: one consists of $^6$Li 
and $^7$Li, and the other consists of $^8$Li, $^9$Li and $^{11}$Li. 
In the former and latter groups we start with 
$^7$Li and $^9$Li, respectively. The energy differences between 
the first $1/2^-$ ($1p_{1/2}$) and ground states ($1p_{3/2}$)
in $^7$Li and $^7$Be are less than 500 keV, while
those in $^9$Li and $^9$Be are less than 3 MeV (see Table~\ref{tab:exc}). 
We therefore assume that the spin-orbit force in $^{7,9}$Li is  
not so strong and that it does not affect much the rms radius and 
single-particle energy levels. First, for each of $^{7,9}$Li 
we roughly determine four potential parameters, 
$r_0$, $a$ and $U_0$, for $n$ and $p$, so as to produce 
the rms radius, $r_A$, as well as 
the one-neutron and one-proton separation 
energies, $S_n$ and $S_p$, 
with a constraint that the four parameters obtained are not far from the 
standard values of the potential parameter set given in Ref.~\cite{BM2}.  
(Note that the two geometrical parameters, $r_0$ and $a$, are 
common to both $p$ and $n$.) 

We next added the spin-orbit potentials, $U_{LS}$, for $n$ and $p$ 
to  fit approximately the energy difference, $\Delta \varepsilon$, between 
the first excited and ground states.  Then, we finally tune  
all the potential parameters to exactly reproduce $S_n$, $S_p$, 
$\Delta \varepsilon(p)$, $\Delta \varepsilon(n)$ and $r_A$  
in the vicinity of the values determined in the rough fitting.  For the 
other nuclei in each group, we start parameter-fitting using 
the values for $^7$Li or $^9$Li 
and find an appropriate set under the condition that the parameter values 
differ from those for $^7$Li or $^9$Li as little as possible.

We present the optimum parameter set in Table~\ref{tab:param} and 
show how the present model reproduces the experimental data 
in Table~\ref{tab:fit}.  
We have also checked numerically how the properties of the Li isotopes depend
on the choice of parameter values.  It is then found that the
nuclear properties depend on $r_0$ and $U_0$ somewhat
strongly, but that its dependence on $a$ and $U_{LS}$ are weak as
we expected first. 

We comment here on the one-neutron separation energy of $^{11}$Li in 
Table~\ref{tab:param}. 
The value of $S_n$ obtained from the experimental binding
energy is 0.326 $\pm$ 0.042 MeV, which is larger than its two-neutron
separation energy, 0.301 $\pm$ 0.029 MeV (see Table~\ref{tab:exp}).
Therefore, the probability for two-neutron
removal may be larger than that for removing one-neutron. 
(Note that $^{10}$Li is observed experimentally as a resonance state.)
Thus, in the present calculation we suppose 
two possibilities for the one-neutron 
separation energy of $^{11}$Li in order 
to see the dependence of the structure 
functions on $S_n$.  We take $S_n$ = 301 or 350 keV, and 
hereafter we call the case of $S_n$ = 301 (350) keV $^{11}$Li-a(b), 
respectively.

As the neutron number increases the central potential for 
neutrons becomes shallow, while that for protons becomes deep (although 
they show an odd-even mass number effect).  This is vital to 
fit the small separation energy of the valence neutrons as well as 
the observed rms radii of Li isotopes.  Remarkably, the strength of 
the spin-orbit force for both protons and neutrons 
grows drastically with increasing $A$. 
In Table~\ref{tab:energy} we present the energy levels of 
the single-particle states in the Li isotopes. The rms radius calculated  
for each single-particle wave function is also given. 
According to the weak binding of the valence neutrons in Li, 
the rms radius of the last neutron becomes large with growing $A$. 

Here we should point out some caveats for the present model (see
Table~\ref{tab:exp}):
(i) In the observed data
the two-neutron separation energy, $S_{2n}$, is not twice
the one-neutron separation energy, while the present calculation
gives $S_{2n} = 2 S_n$. (ii) In the low-lying spectra
of $^6$Li and $^8$Li their degeneracies are decoupled in the experiments.
(iii) The distributions of protons and neutrons in $^{8,9}$Li are quite
different at large $r$ in the calculation 
(see also subsection~\ref{subsec:kinetic}). 
Those difficulties indicate the necessity of including
residual two-body interactions in nuclei.
The first defect could be amended by considering the effect of a
pairing force between neutrons, while the last one may indicate that a
$p$-$n$ pairing is required to suppress the leakage of
neutrons in $^{8,9}$Li.
Note that neutron leakage or shrunkage of the proton distribution influences
the nuclear structure functions much less than the separation energy. 
This fact allows us to use the present parameter sets for their
single-particle potentials, where $r_0$ and $a$ are common
to both $p$ and $n$, in spite of the caveat (iii). Furthermore, since 
we used effective potentials which reproduce 
the separation energies of nucleons close to the Fermi surface, 
and it is hard to measure experimentally the 
separation energies of deeply bound nucleons, it may not be necessary to 
be too concerned about the caveats (i) and (ii) in the present calculation. 

\subsection{Nucleon kinetic energy and average binding energy}
\label{subsec:kinetic}

\begin{figure}
\begin{center}
\epsfig{file=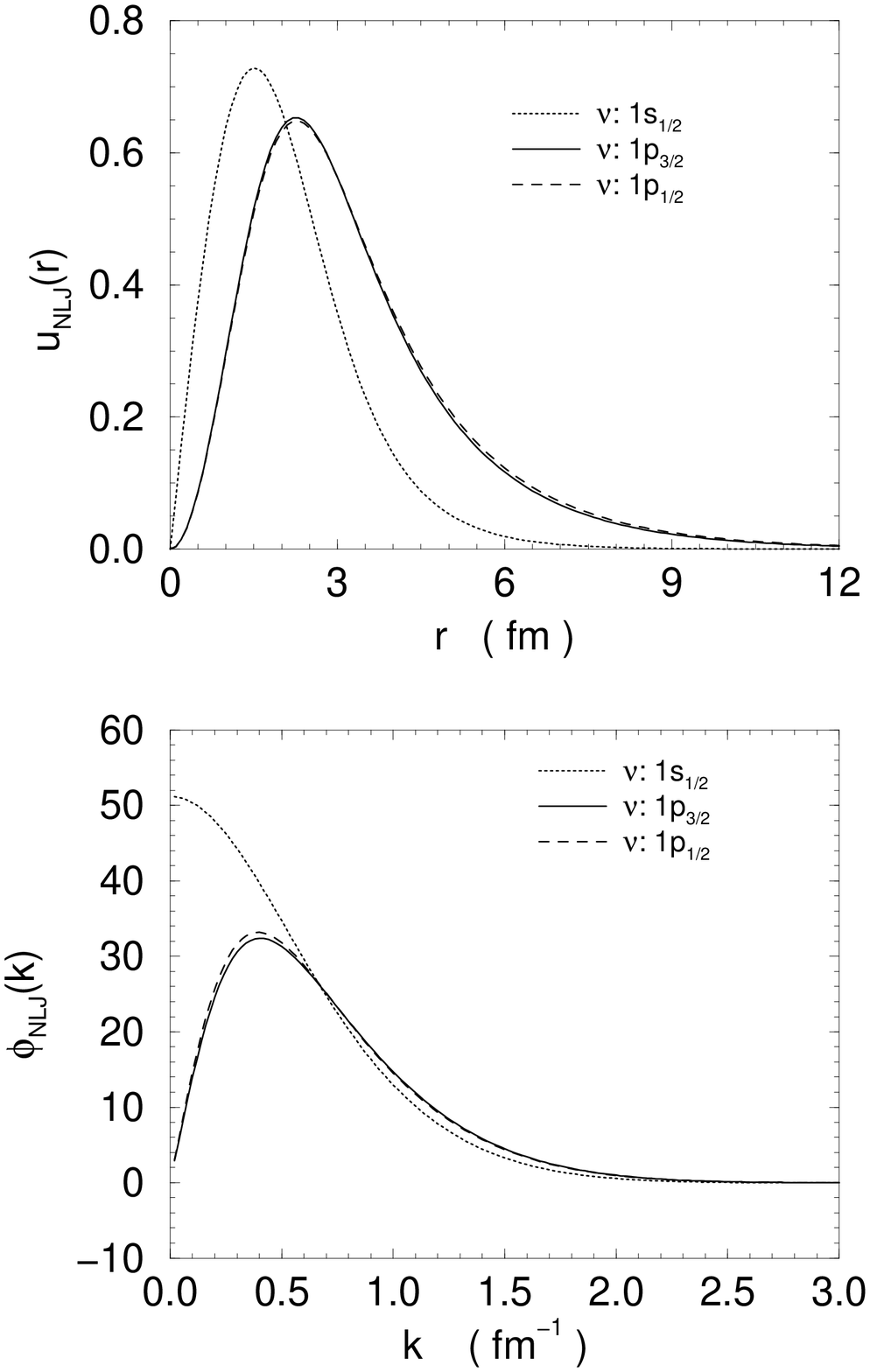,height=10cm}
\epsfig{file=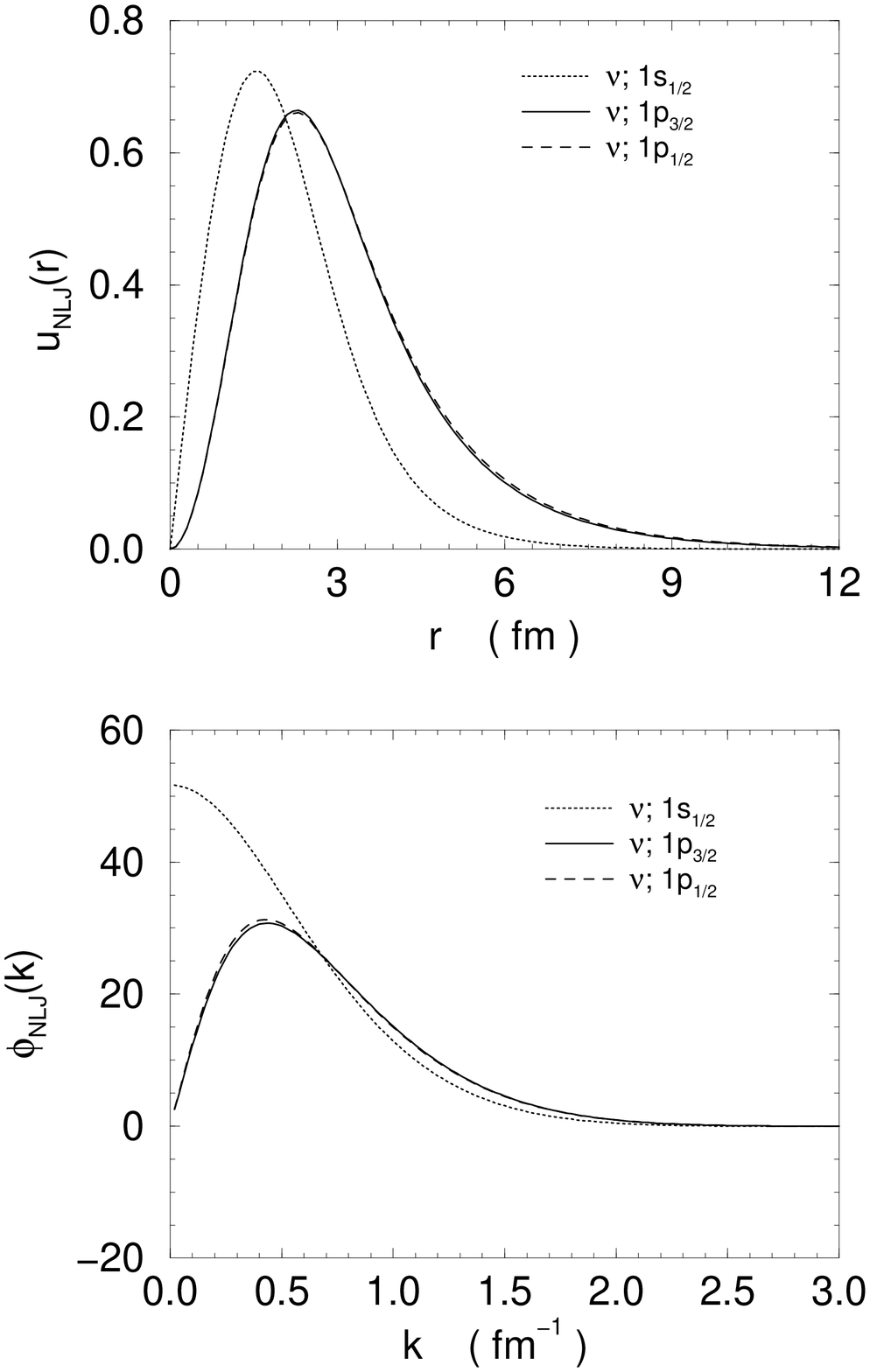,height=10cm}
\epsfig{file=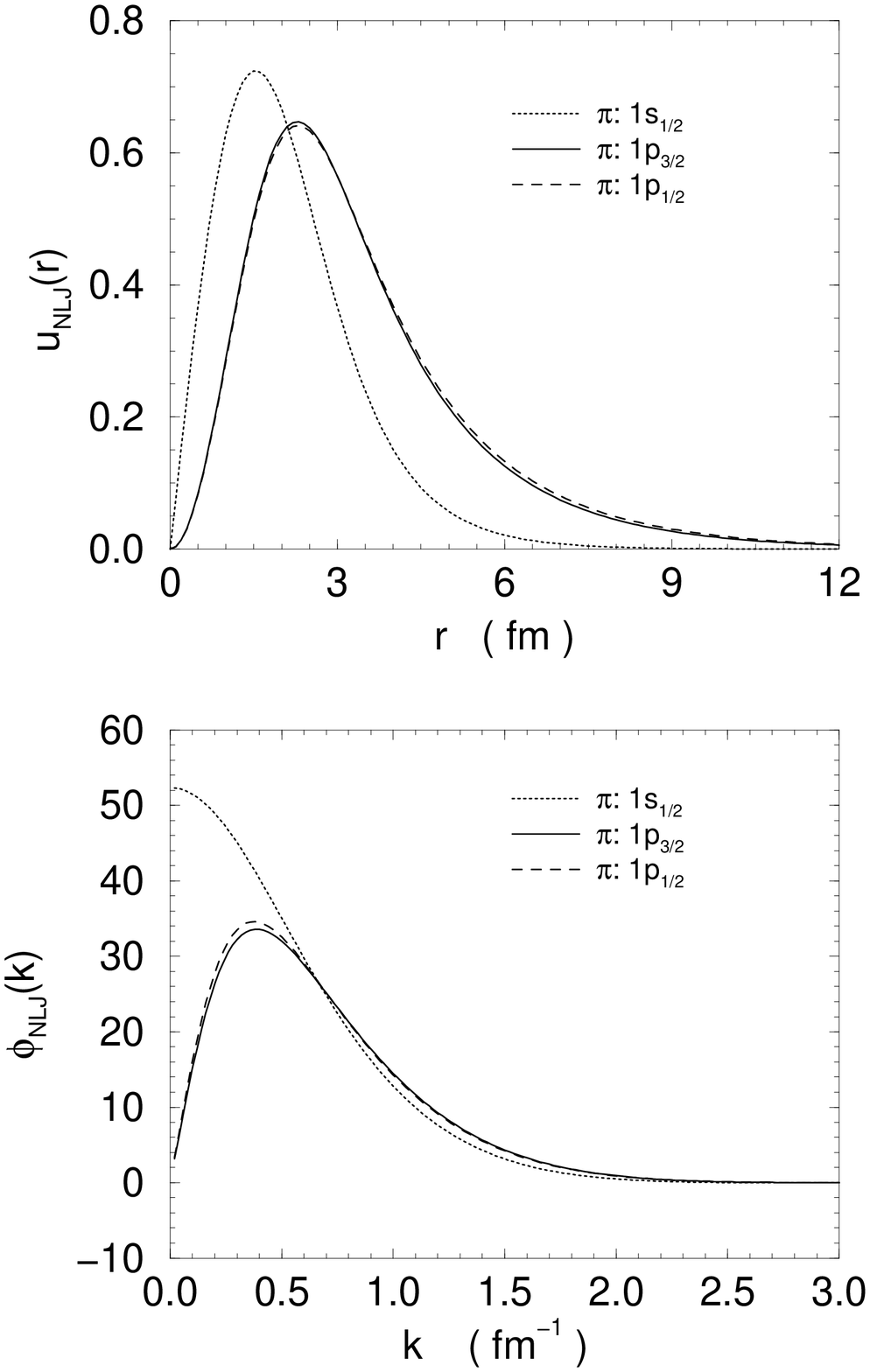,height=10cm}
\epsfig{file=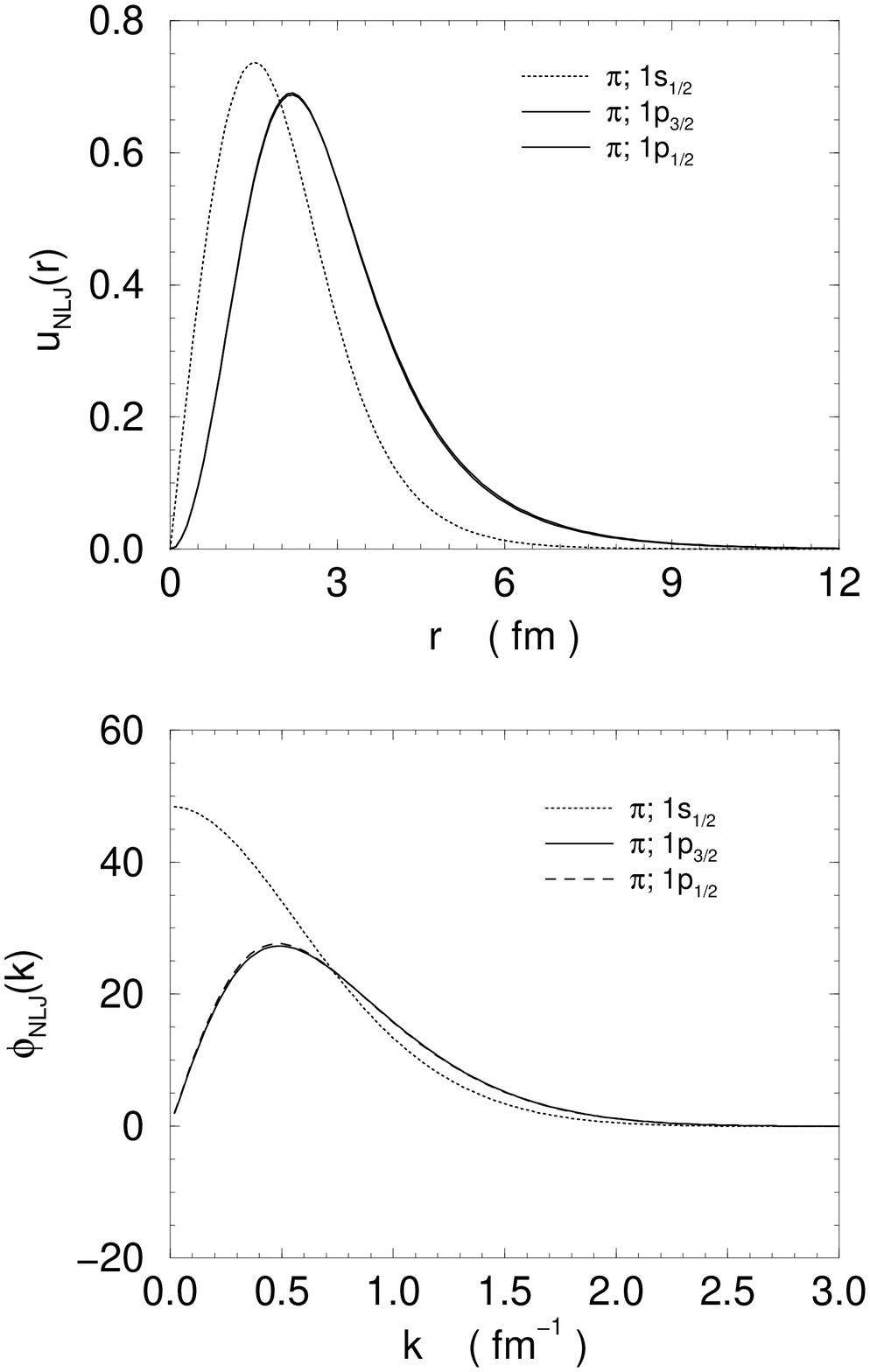,height=10cm}
\caption{Radial wave functions, $u_{\alpha}(r)$, and momentum distributions,
$\phi_{\alpha}(k)$, of the single-particle states in $^6$Li (left panels)
and $^7$Li (right panels). The top two panels are for neutrons, while
the bottom two are for protons.
In each panel the dotted, solid and dashed lines denote the distributions
of the nucleons in the 1$s_{1/2}$, 1$p_{3/2}$ and 1$p_{1/2}$ shell orbits,
respectively. }
\label{fig:mom67}
\end{center}
\end{figure}

\begin{figure}
\begin{center}
\epsfig{file=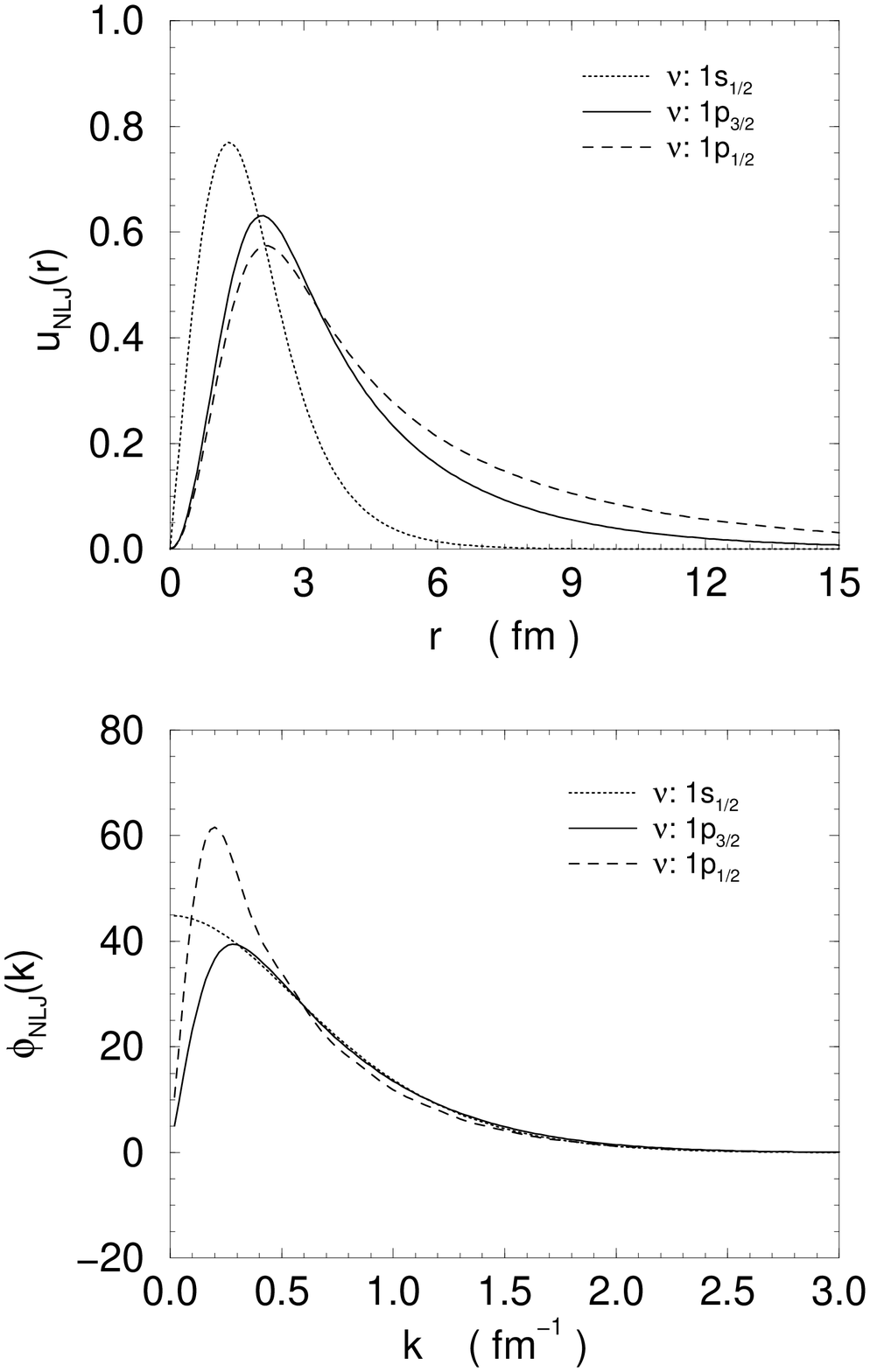,height=10cm}
\epsfig{file=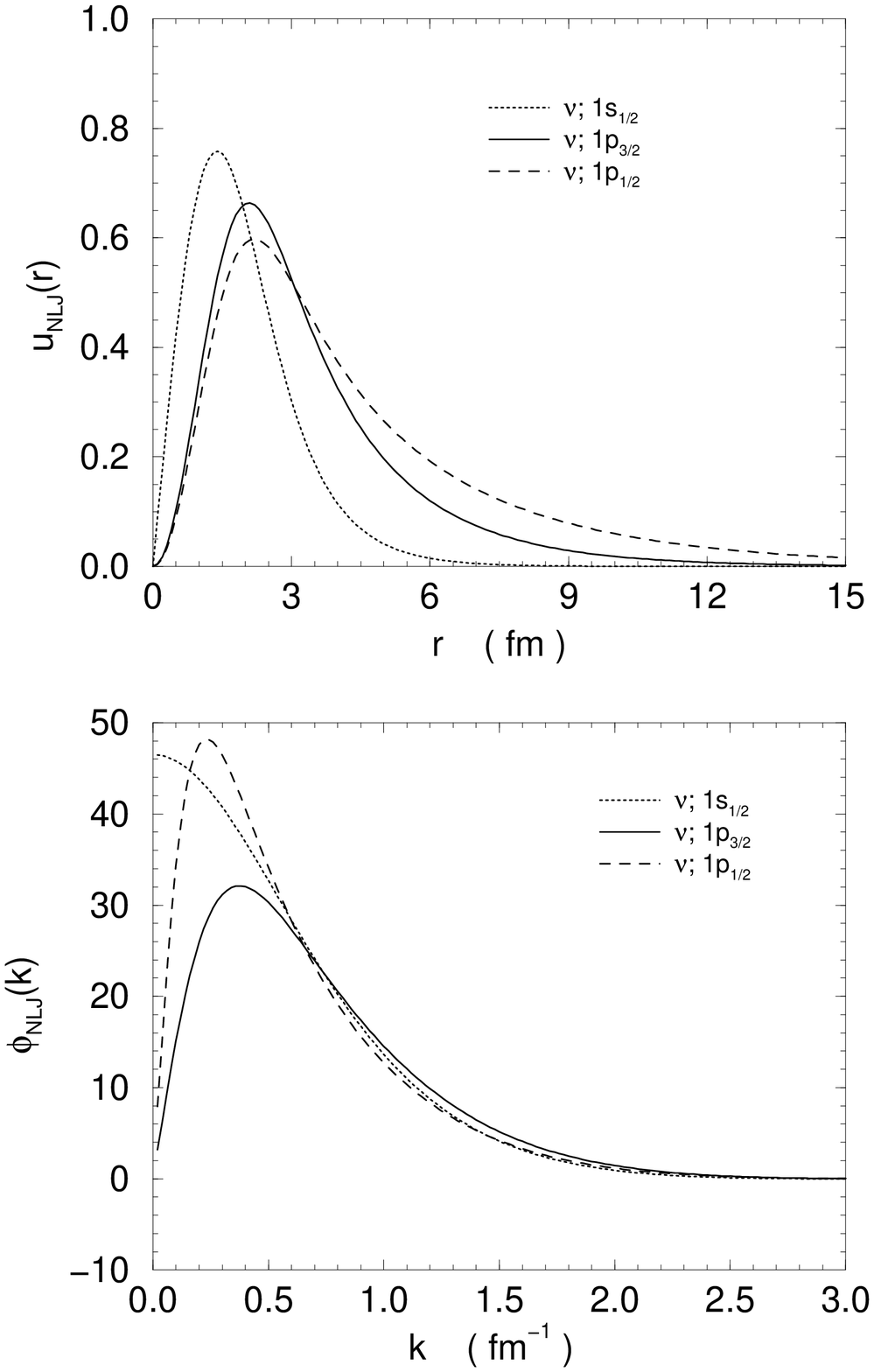,height=10cm}
\epsfig{file=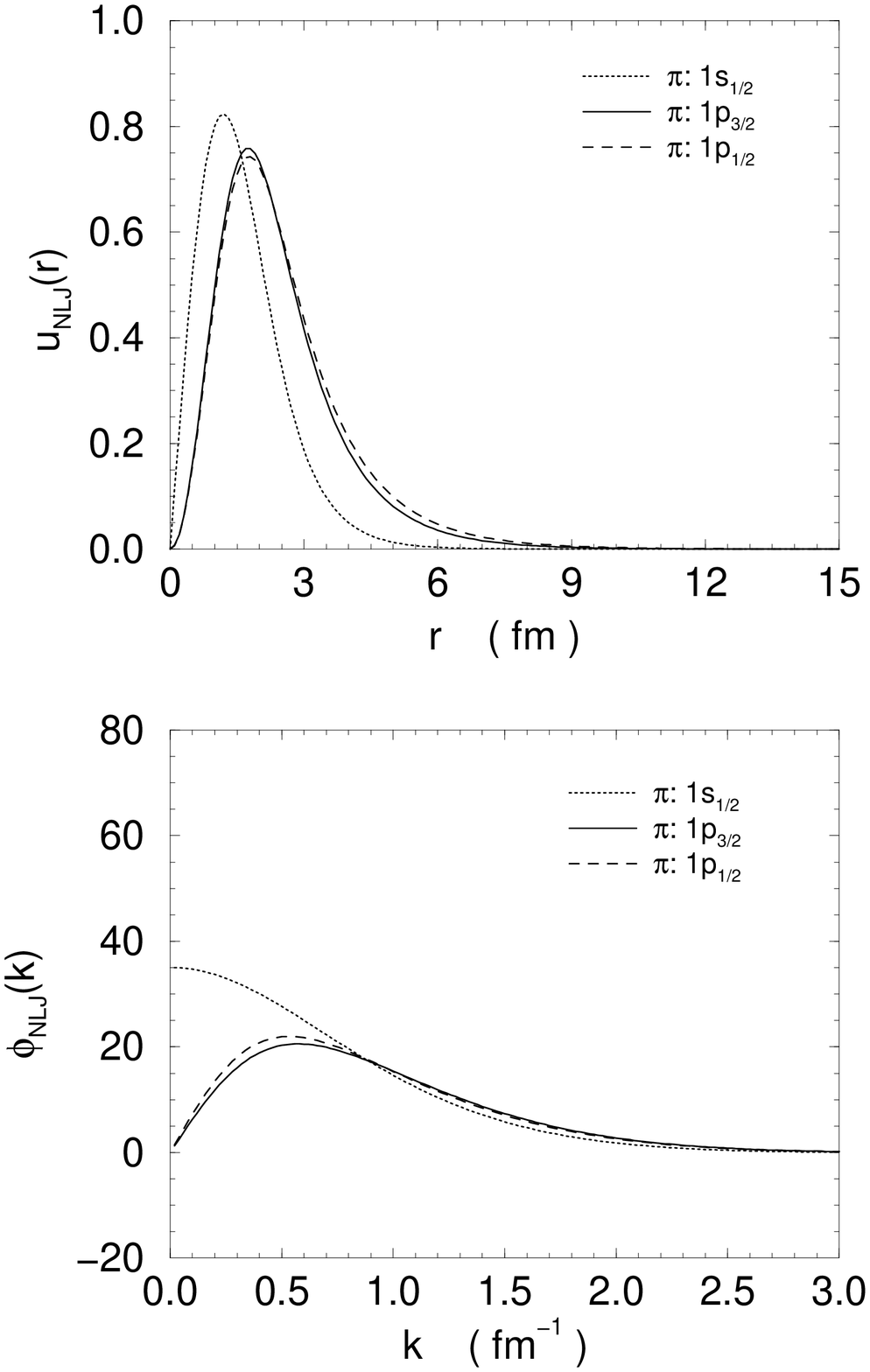,height=10cm}
\epsfig{file=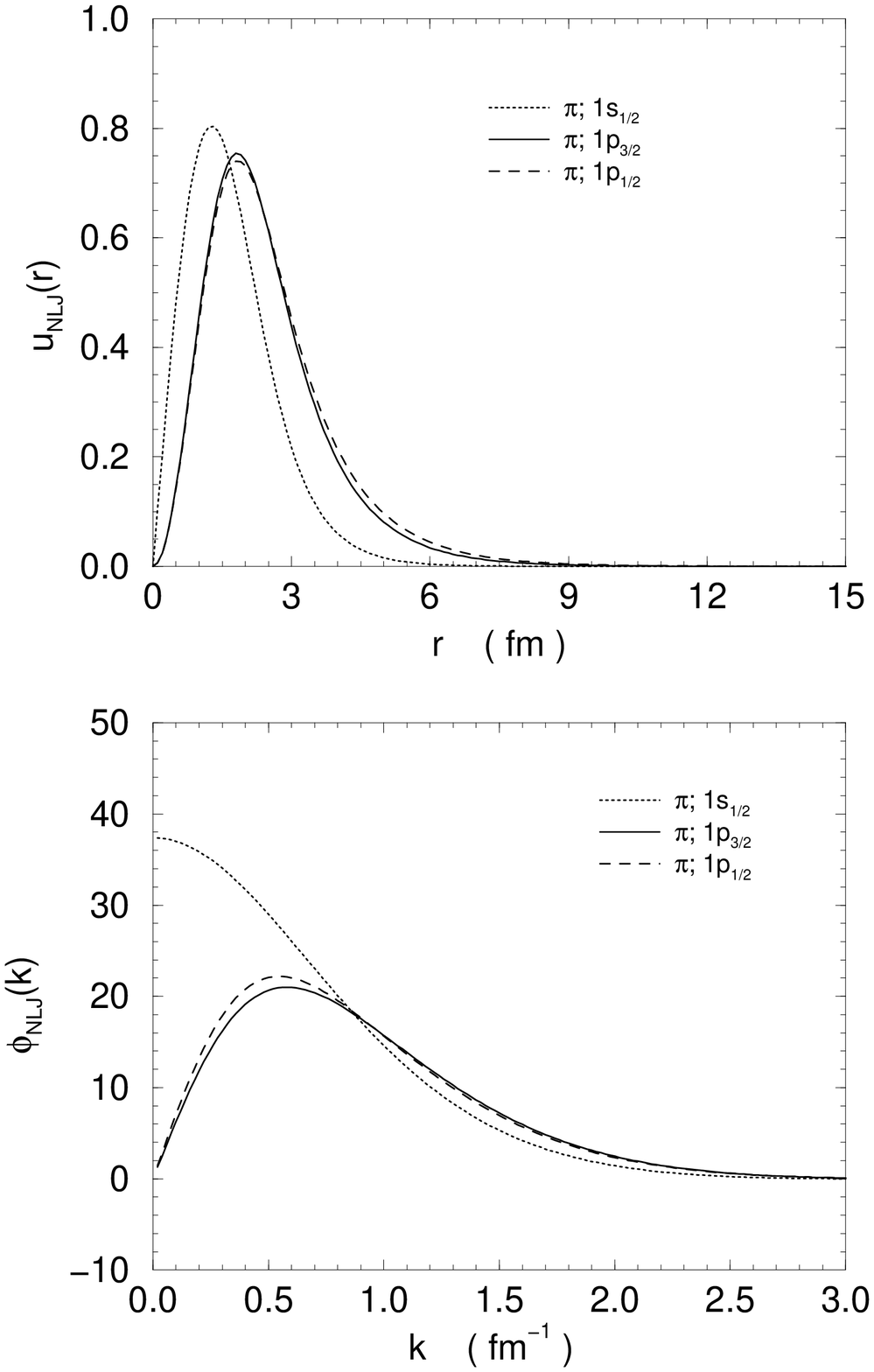,height=10cm}
\caption{Same as Fig.~\protect\ref{fig:mom67} 
but for $^8$Li (left panels) and $^9$Li (right panels).}
\label{fig:mom89}
\end{center}
\end{figure}

\begin{figure}
\begin{center}
\epsfig{file=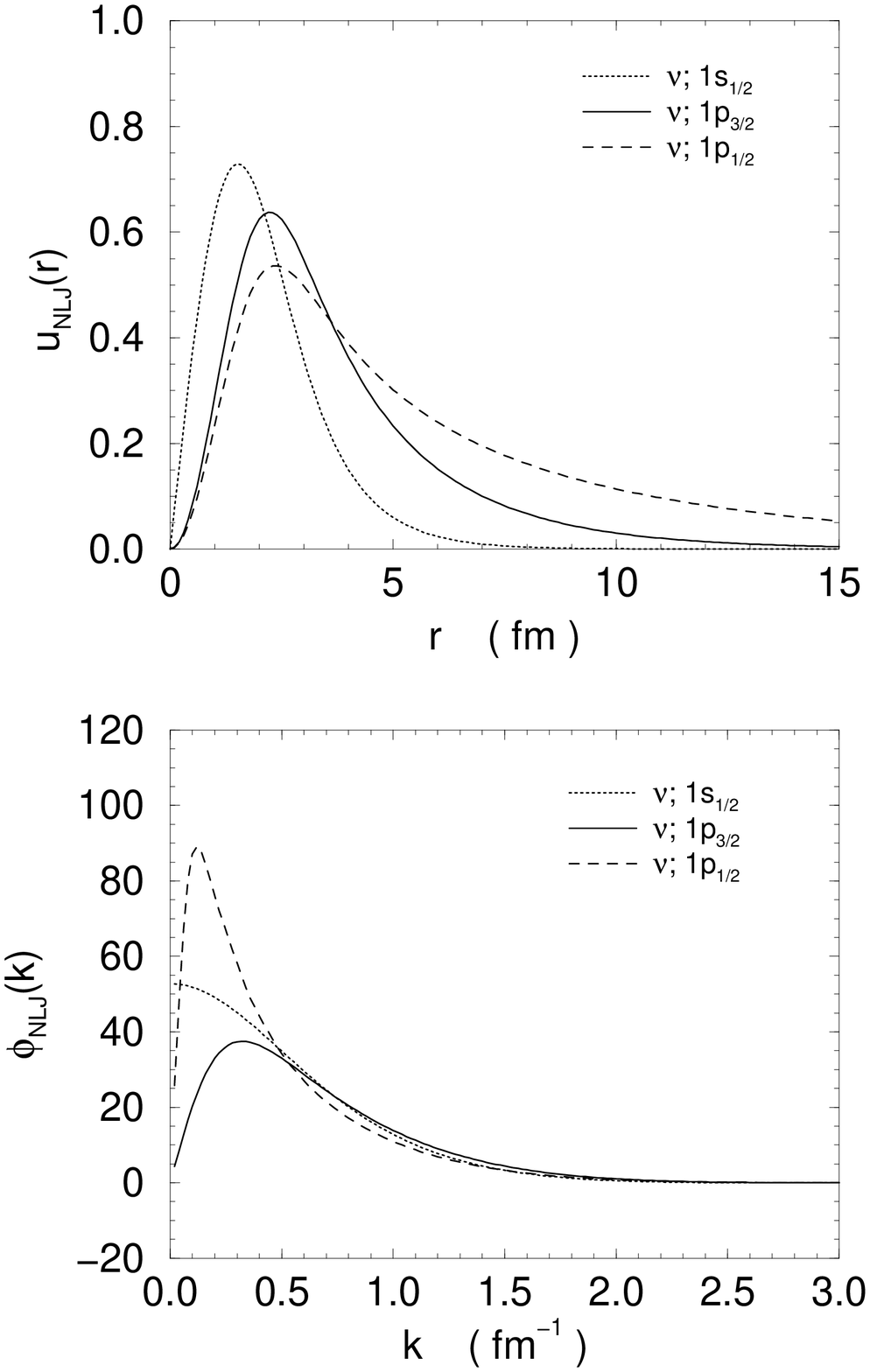,height=10cm}
\epsfig{file=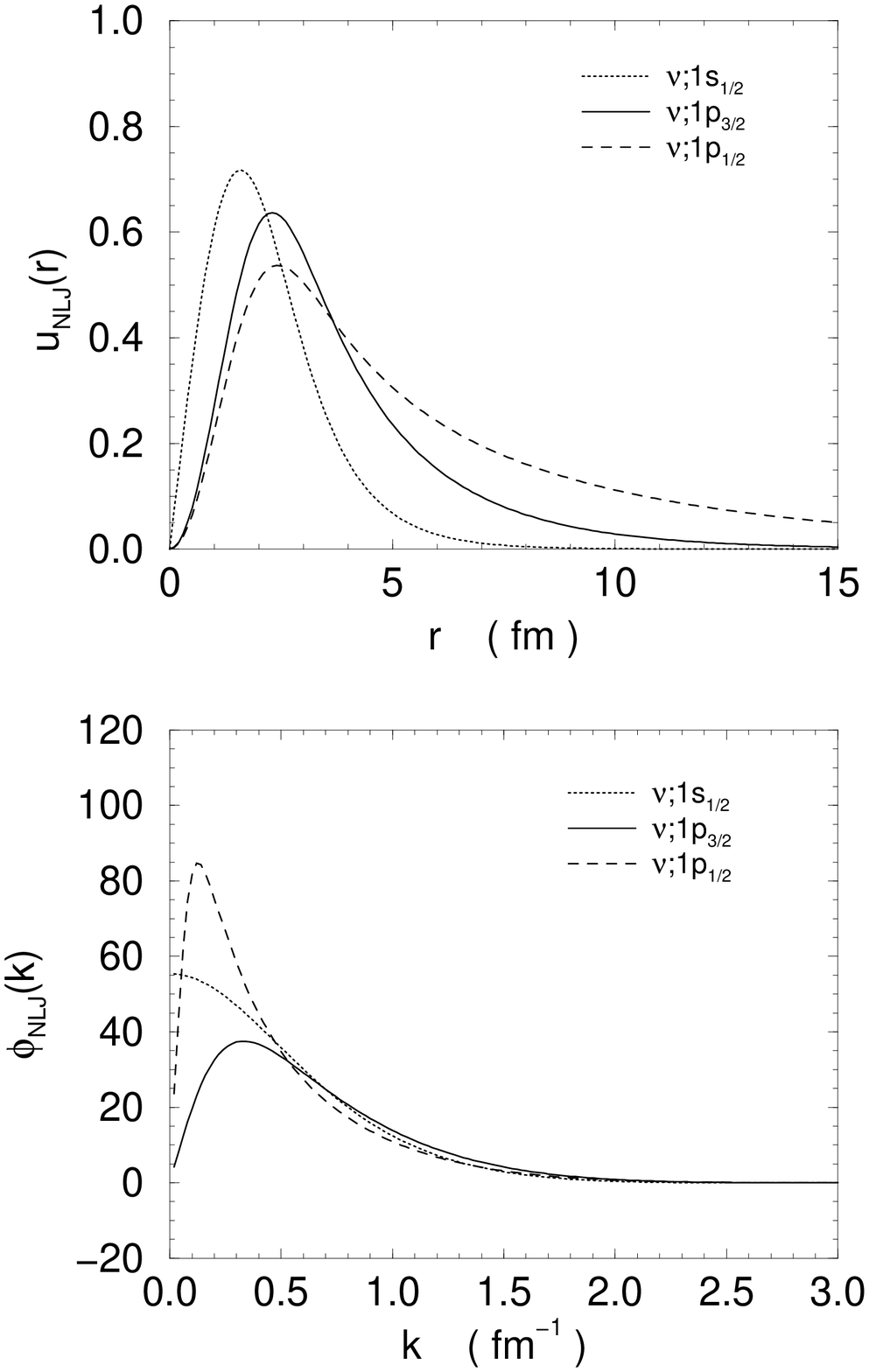,height=10cm}
\epsfig{file=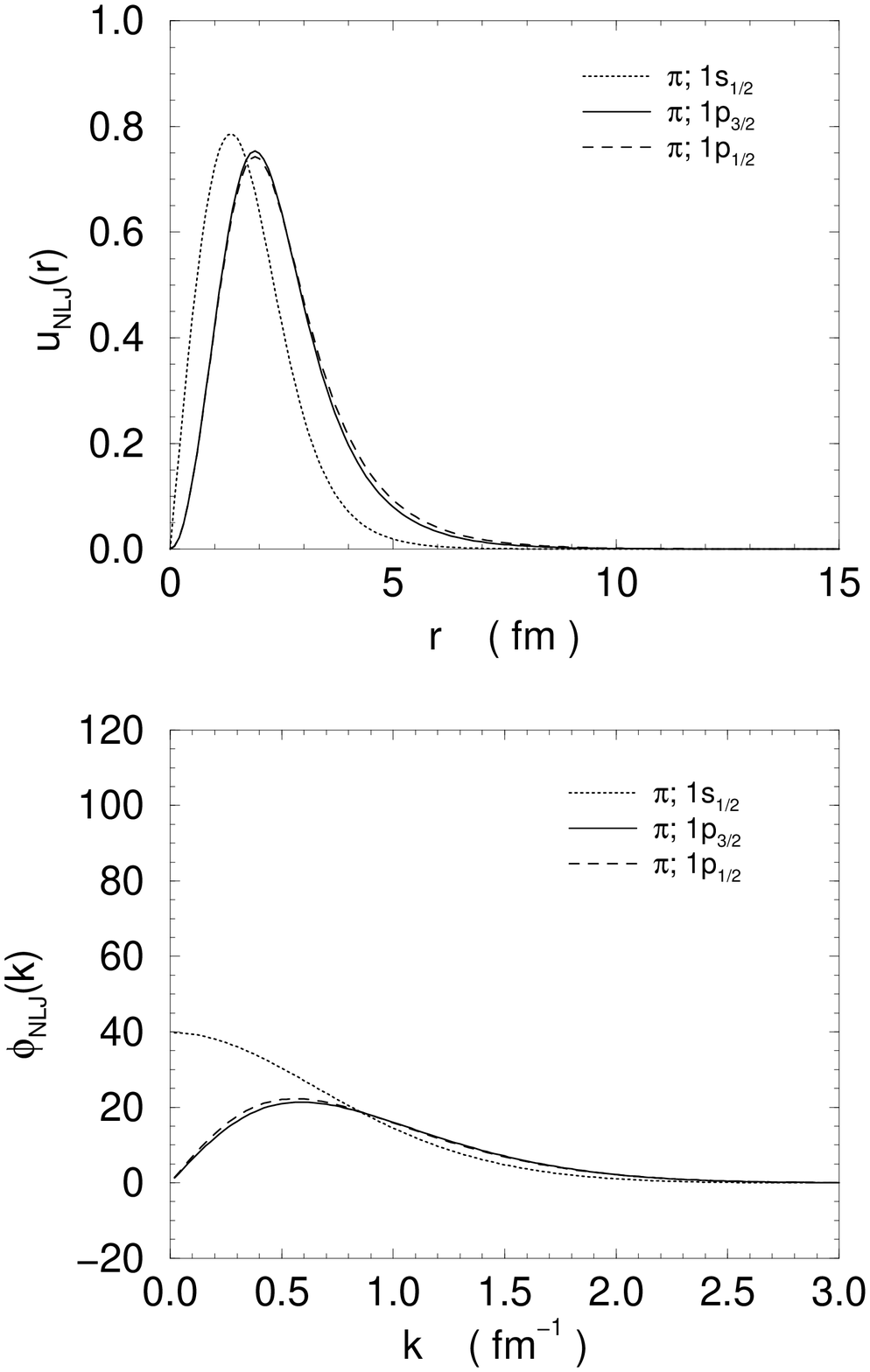,height=10cm}
\epsfig{file=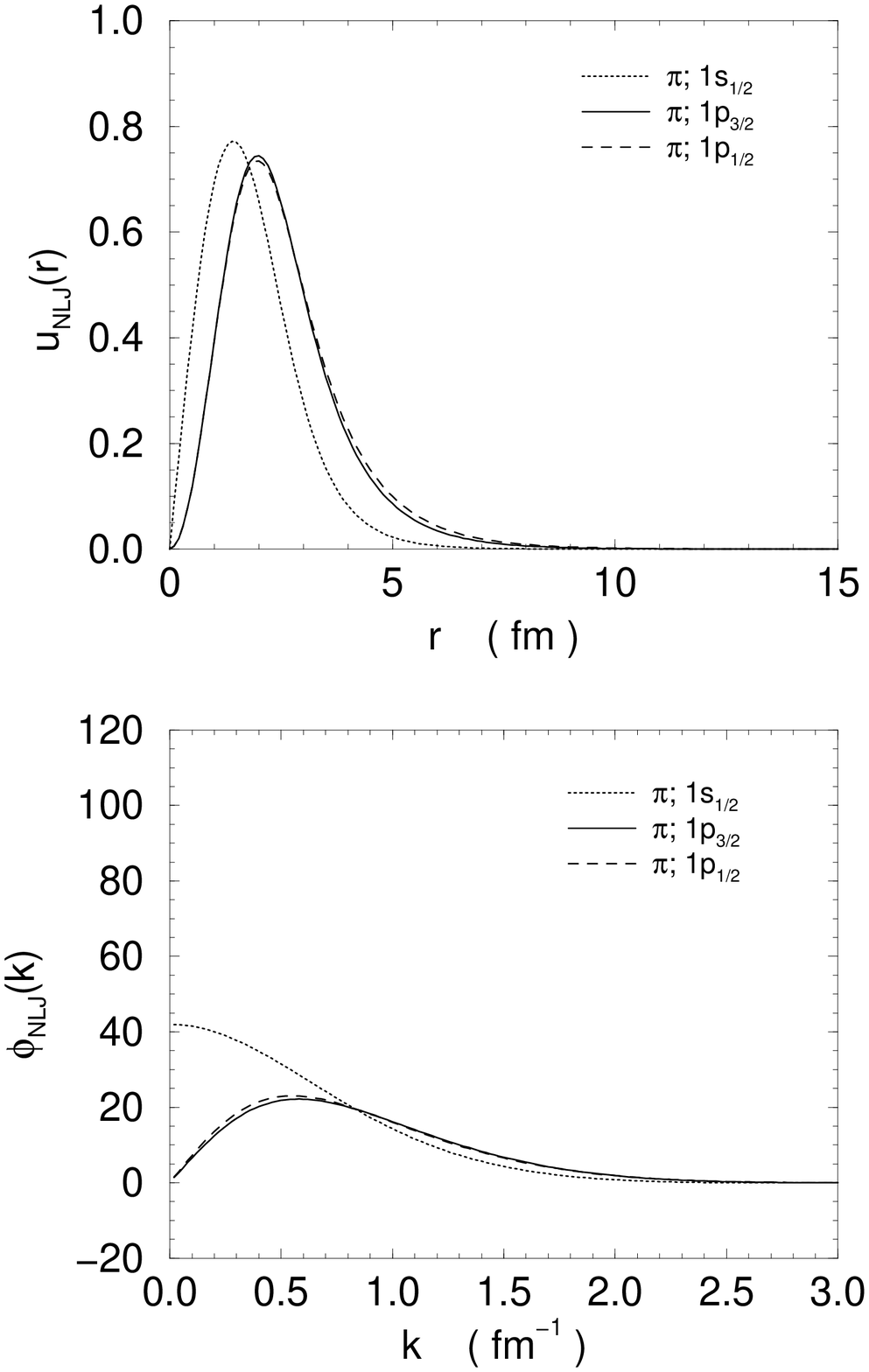,height=10cm}
\caption{Same as Fig.~\protect\ref{fig:mom67} but for 
$^{11}$Li. The left 
and right panels are the results for the cases of 
$S_n = 301$ keV and 350 keV, respectively.} 
\label{fig:mom11}
\end{center}
\end{figure}

\begin{figure}
\begin{center}
\epsfig{file=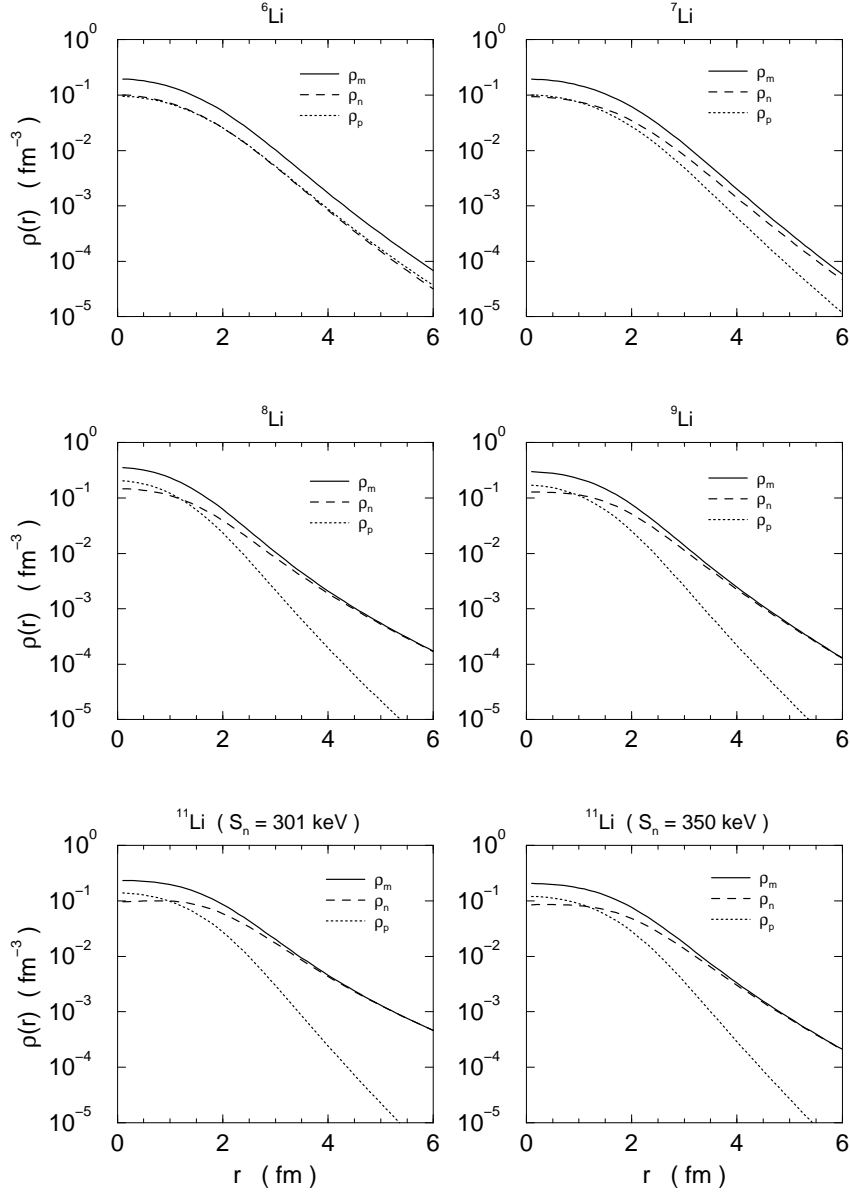,height=16cm}
\caption{Matter ($\rho_m$), neutron ($\rho_n$) and proton ($\rho_p$) 
density distributions in 
$^{6-11}$Li. The neutron- and proton density distributions
are obtained by summing up of the square of the corresponding single-particle 
wave functions. The matter distribution is simply given by $\rho_p + \rho_n$. 
} 
\label{fig:density}
\end{center}
\end{figure}

In Figs.~\ref{fig:mom67} - \ref{fig:mom11} we present the wave 
functions in coordinate space, $u_{\alpha}(r)$, 
and the radial momentum distribution of a nucleon, 
$\phi_{\alpha}(k)$, in the Li isotopes.
One can see that as the neutron number increases the tail of the neutron 
wave function in the $1p$ orbit is longer.  Accordingly,
the distribution of the $1p$ neutron becomes sharp in momentum space.
As expected, the valence neutrons in the halo of $^{11}$Li
distribute very broadly in coordinate space and their
momentum distributions are quite sharp around $k \sim 0.1$ fm$^{-1}$.
The density distributions are illustrated in Fig.~\ref{fig:density}.
As pointed out in subsection~\ref{subsec:comments}, the neutrons leak
out to large $r$ even in $^{8,9}$Li, which may imply
that a proper (attractive) $p$-$n$ pairing force is required.

In an unpolarized nucleus the kinetic energy of a nucleon in the 
single-particle shell, $\alpha = \{n, \ell, j, \tau_z \}$, 
can be calculated by
\begin{equation}
\langle t_{\alpha} \rangle = 
\frac{1}{2j+1} \sum_{m} \int \frac{d^3k}{(2 \pi)^3}
\vert \phi_{n \ell j m \tau_z}({\vec k}) 
\vert^2 \, \frac{k^2}{2M} = 
\frac{1}{2M} \int_0^{\infty} \frac{k^4 dk}{(2 \pi)^3}
\vert \phi_{\alpha=\{n \ell j \tau_z\}}(k) 
\vert^2  .
\label{nkinetic}
\end{equation}
Here $\phi_{n \ell j m \tau_z}({\vec k})$ is the nucleon momentum 
distribution, which is given by Fourier transformation of the wave 
function in coordinate space, and it is normalized as 
\begin{equation}
\frac{1}{(2 \pi)^3} \int d^3k \, 
\vert \phi_{n \ell j m \tau_z}({\vec k}) \vert^2 = 1 .
\label{norm}
\end{equation}

The kinetic energy of the center-of-mass (cm) motion of the nucleus
is then evaluated by
\begin{equation}
T_{\rm cm} = \frac{1}{A} \sum_{i} \langle t_{\alpha_i} \rangle , 
\label{cm}
\end{equation}
and the spurious cm motion should be subtracted from 
the total energy. We thus obtain the mean-nucleon 
kinetic energy, $\langle t \rangle$, in the nucleus 
and the total binding energy: 
\begin{eqnarray}
\langle t \rangle &=& \frac{1}{A} 
\biggl ( 1 -\frac{1}{A} \biggr ) \sum_{i}
\langle t_{\alpha_i} \rangle  , 
\label{average} \\
\langle B \rangle &=& 
\frac{1}{2} \sum_{i} (\varepsilon_{i}+\langle t_{\alpha_i} \rangle) 
-T_{\rm cm} , 
\label{totenergy}
\end{eqnarray}
respectively.
In Table~\ref{tab:kien} we summarize the kinetic 
energy of the nucleon in each shell model orbit. 
We should note that the calculated binding energies are rather smaller 
than the observed ones (see Table~\ref{tab:exp}).  With increasing the  
neutron number the total binding energy decreases.  Probably this 
discrepancy may be improved by taking residual two-body correlations
into account. However, we again emphasize that the present calculation
reproduces the observed separation energies of Li isotopes, which plays the
most important role in the calculation of nuclear structure functions. 

\section{Spin-average structure functions of Li isotopes}
\label{sec:Listrfn}
\subsection{Light-cone momentum distributions and nuclear structure functions}
\label{subsec:momenta}

In the DIS of high energy 
charged leptons off an unpolarized  nucleus one can observe the distribution 
of the {\em longitudinal} momentum fraction, $x$, carried by the 
quarks in the nucleus~\cite{review}. The nuclear structure function, 
$F_2^A(x)$, is usually described by a convolution of momentum distributions 
of nucleons in $A$ and the free nucleon structure function~\cite{review,pw}:
\begin{equation}
F_2^A(x) = \sum_{i}^{\rm A} \int_x^A dy \, f_{\alpha_i/A}(y) F_2^i(x/y) , 
\label{conv}
\end{equation}
where $F_2^i = F_2^{p(n)}$ when the $i$-th nucleon is a proton (neutron). 
(Note that in the scaling limit $F_2^A(x)$ is related to $F_1^A(x)$ 
through the Callan-Gross relation. See also Eq. (\ref{relation}).) 
$f_{\alpha_i/A}(y)$ is a distribution of the 
{\it longitudinal} momentum fraction, $y$, carried by the nucleon 
in the nucleus $A$, which is defined as 
\begin{equation}
0 \le y/A \equiv \frac{k \cdot q}{P \cdot q} \le 1 
\hspace{1cm} (0 \leq y \leq A) ,
\label{ydef}
\end{equation}
with $P^\mu$ the four momentum of the nucleus $A$. 

In the scaling limit the light-cone momentum distribution in 
the target rest frame is then written as~\cite{review,pw}
\begin{equation}
f_{\alpha_i/A}(y) = \int \frac{d^4k}{(2 \pi)^4} \, S_{\alpha_i/A}(k)
\biggl ( 1 + \frac{k^3}{k^0} \biggr ) \delta \biggl (
y - \frac{k^0+k^3}{M} \biggr )  , 
\label{fiy}
\end{equation}
with the nucleon spectral function 
\begin{equation}
S_{\alpha_i/A}(k) = 2\pi \delta(k_0-M-\varepsilon_i+T_R)
\vert \Psi_{\alpha_i}({\vec k}) \vert^2 ,
\label{S}
\end{equation}
where $-\varepsilon_i (>0)$ is the separation energy of the $i$-th nucleon 
in the present approximation, 
$T_R = {\vec k}^2/2M_{A-1} (> 0)$ the recoil energy
of the residual nucleus $(A-1)$ with its mass $M_{A-1}$, and
$\Psi_{\alpha_i}({\vec k})$ the three-momentum distribution
of the struck nucleon $i$.  For an unpolarized target, 
$\vert \Psi_{\alpha_i}({\vec k}) \vert^2$ is given in terms of the 
momentum distribution of the nucleon as
\begin{equation}
\vert \Psi_{\alpha_i}({\vec k}) \vert^2 \equiv \frac{1}{2j +1} \sum_m 
\vert \phi_{n \ell j m \tau_z}({\vec k}) \vert^2 
= \frac{1}{4\pi} \vert \phi_{\alpha_i}(k) \vert^2 , 
\label{Psi}
\end{equation}
which is spherically symmetric and independent of the angular variables. 

Inserting Eq. (\ref{S}) into Eq. (\ref{fiy}), we find  
\begin{equation}
f_{\alpha_i/A}(y) = \int \frac{d^4k}{(2 \pi)^4}
2 \pi \delta(k_0-M-\varepsilon_i+T_R)
\vert \Psi_{\alpha_i}({\vec k}) \vert^2
\biggl ( 1 +\frac{k^3}{k^0} \biggr )
\delta\biggl ( y - \frac{k^0+k^3}{M} \biggr ) .
\label{ggg}
\end{equation}
Using a new notation
\begin{equation}
\bar{M} = \bar{M}(k) = M+\varepsilon_i-T_R ,
\label{barm}
\end{equation}
we can rewrite Eq. (\ref{ggg}) as
\begin{eqnarray}
f_{\alpha_i/A}(y) &=& \int \frac{d^4k}{(2 \pi)^4}
2 \pi \delta(k_0-\bar{M})
\vert \Psi_{\alpha_i}({\vec k}) \vert^2
\biggl ( 1 +\frac{k^3}{k^0} \biggr )
\delta\biggl ( y - \frac{k^0+k^3}{M} \biggr ) \nonumber \\
&=& \frac{M}{2(2 \pi)^3} \int_0^{\infty}
k dk \, \vert \phi_{\alpha_i}(k) \vert^2
\int_{-1}^{+1} \, d{\chi} \,
\biggl ( 1 +\frac{k \chi}{\bar{M}} \biggr )
\delta\biggl ( \chi - \frac{M}{k}
\biggl ( y-\frac{\bar{M}}{M} \biggr ) \biggr ) . 
\label{finfy}
\end{eqnarray}
The distribution $f_{\alpha_i/A}(y)$ vanishes unless
\begin{equation}
\biggl \vert \frac{M}{k} \biggl ( y -\frac{\bar{M}}{M}
\biggr )  \biggr \vert < 1  ,
\label{cond0}
\end{equation}
that is, 
\begin{equation}
\vert M y -\bar{M} \vert =
\vert M y -M-\varepsilon_i+T_R \vert  \leq k  .
\label{cond}
\end{equation}
Since the recoil energy, $T_R$, depends on $k^2$, 
Eq. (\ref{cond}) yields a transcendental equation.
However, $T_R$ is practically small and hence we may replace $T_R$
by its expectation value
\begin{equation}
T_R \to \langle T_R \rangle_{\alpha_i} \simeq 
\frac{\langle k^2 \rangle_{\alpha_i}}{2M_{A-1}} 
\simeq \frac{\langle t_{\alpha_i} \rangle}{A-1} . 
\label{TR}
\end{equation}
Defining the lower limit of the $k$-integral in Eq. (\ref{finfy}) as 
\begin{equation}
k_L \equiv \vert My - M - \varepsilon_i +
\langle T_R \rangle_{\alpha_i} \vert , 
\label{llimit}
\end{equation}
we find 
\begin{equation}
f_{\alpha_i/A}(y) \simeq \frac{M^2}{{\bar M}} y 
\int_{k_L}^{\infty} \frac{k \, dk}{2 (2 \pi)^3}
\, \vert \phi_{\alpha_i}(k) \vert^2 . 
\label{fin2fy}
\end{equation}

\begin{figure}
\begin{center}
\epsfig{file=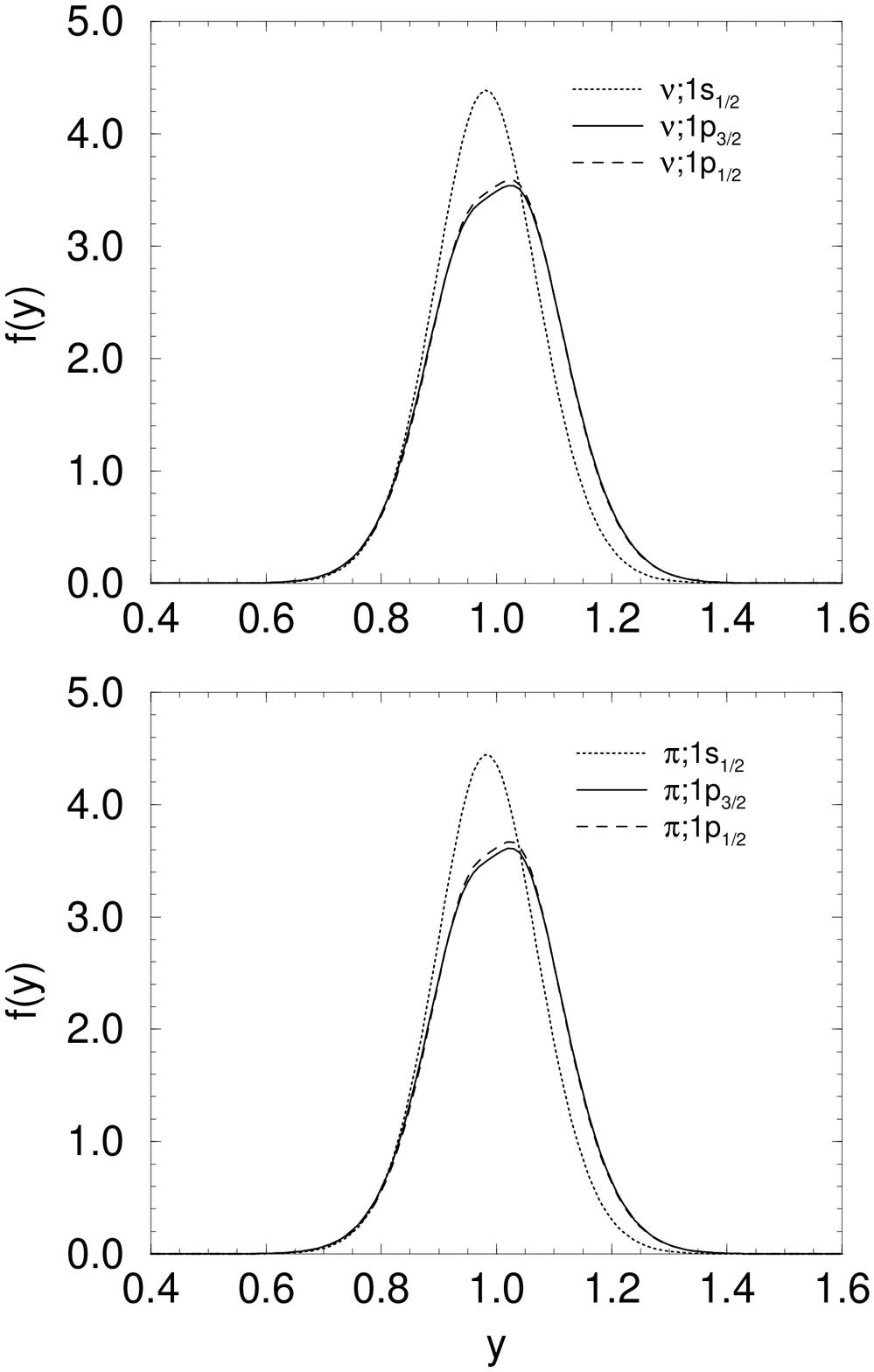,height=10cm}
\epsfig{file=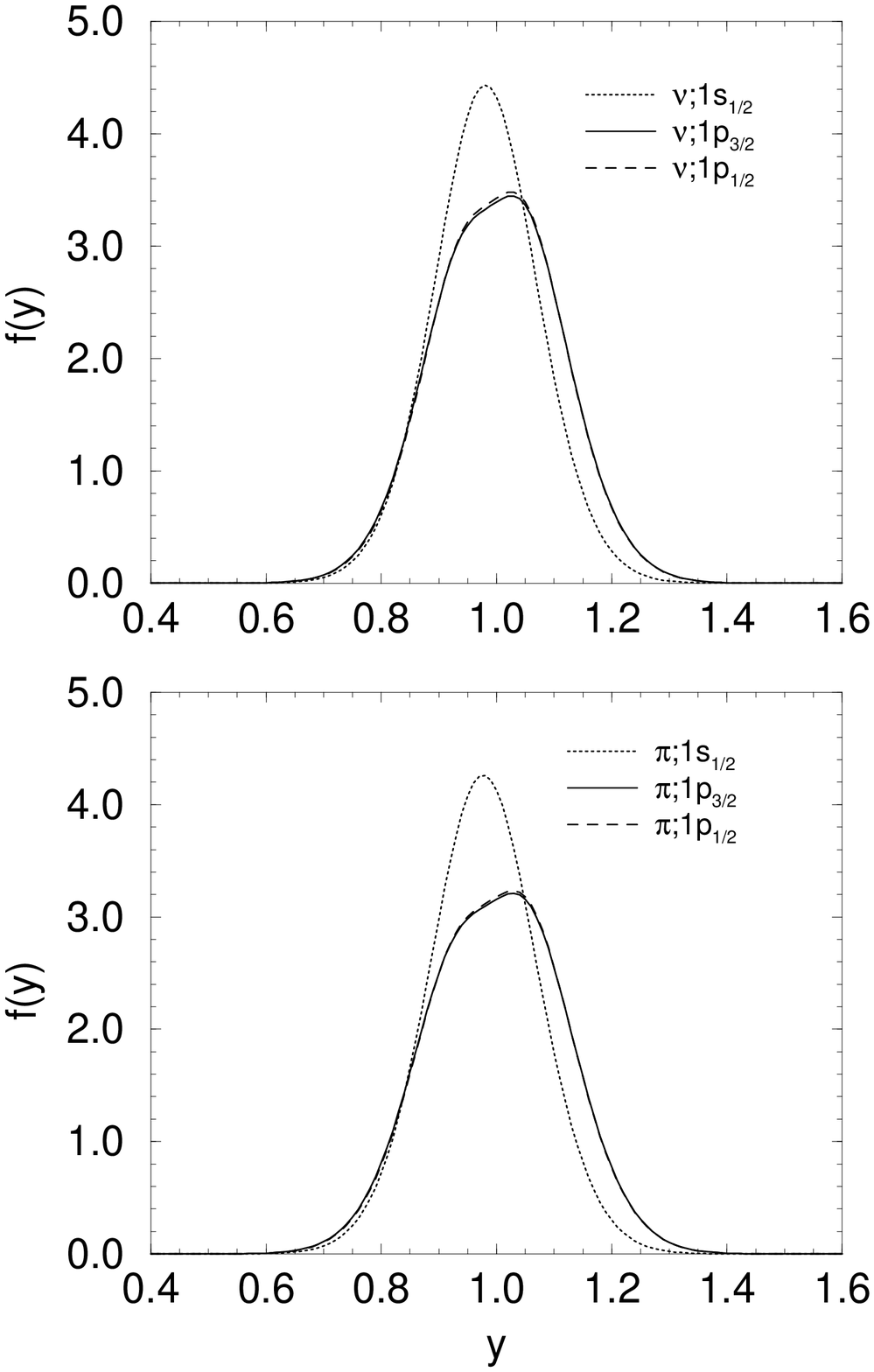,height=10cm}
\caption{$f(y)$ 
for $^{6}$Li (left panels) and $^{7}$Li (right panels). 
In each column the top and bottom panels show those of neutrons and protons, 
respectively. In each panel the dotted, solid and dashed lines denote
the distributions of the nucleons in $1s_{1/2}$, $1p_{3/2}$ and $1p_{1/2}$ 
shell orbits, respectively.}
\label{fig:y67}
\end{center}
\end{figure}

\begin{figure}
\begin{center}
\epsfig{file=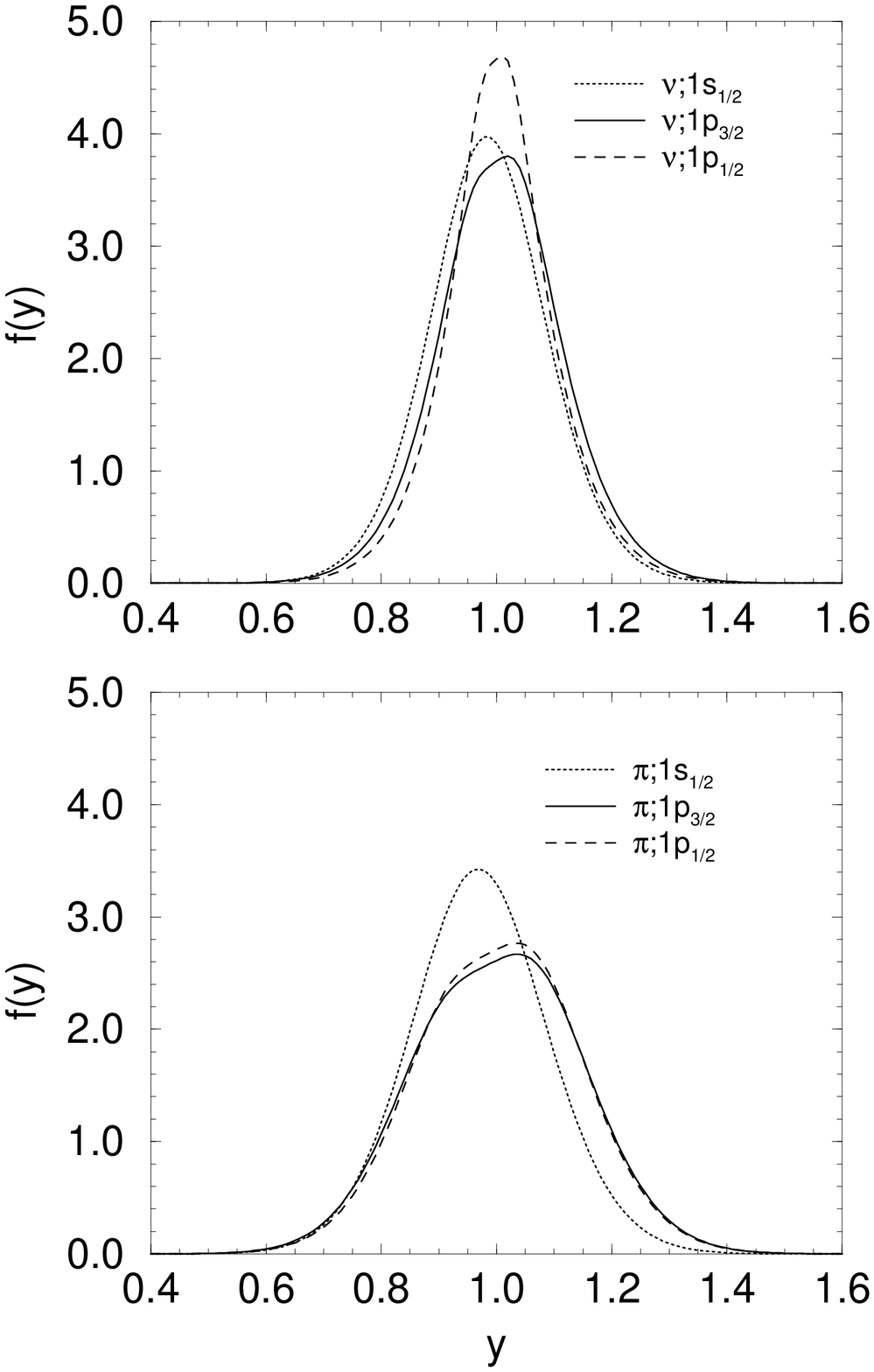,height=10cm}
\epsfig{file=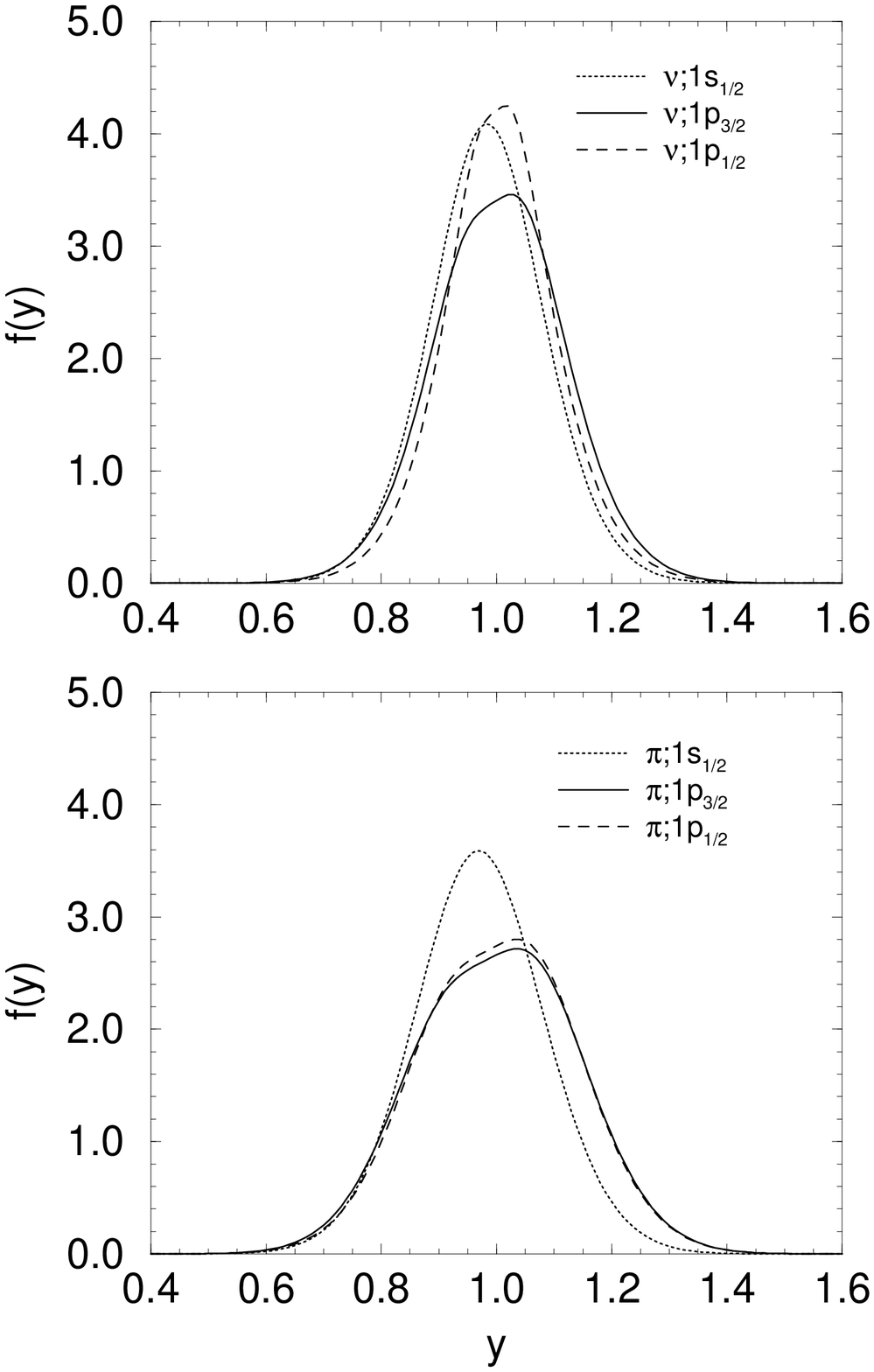,height=10cm}
\caption{Same as Fig.~\protect\ref{fig:y67} but 
for $^{8}$Li (left panels) and $^{9}$Li (right panels).}
\label{fig:y89}
\end{center}
\end{figure}

\begin{figure}
\begin{center}
\epsfig{file=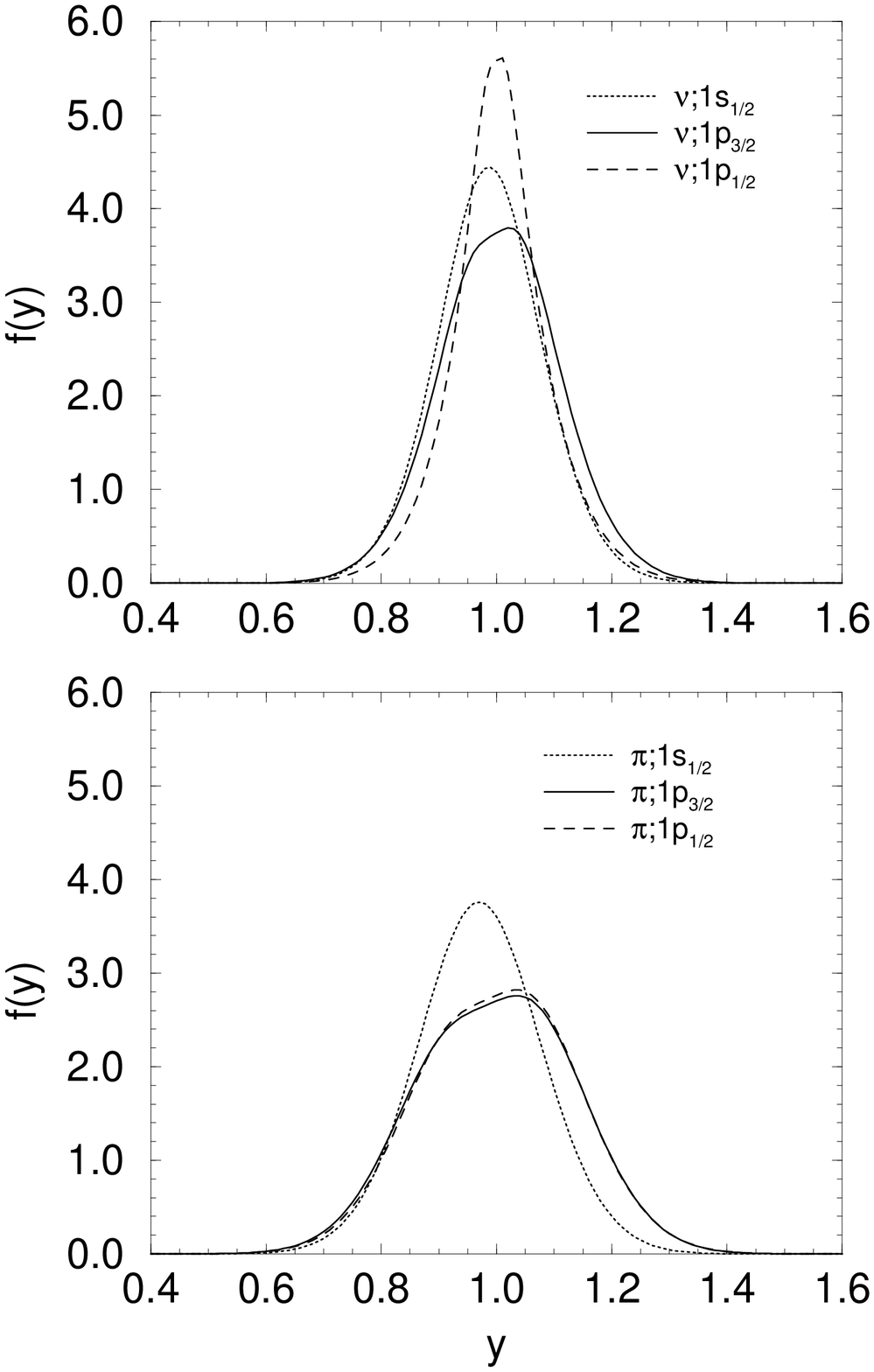,height=10cm}
\epsfig{file=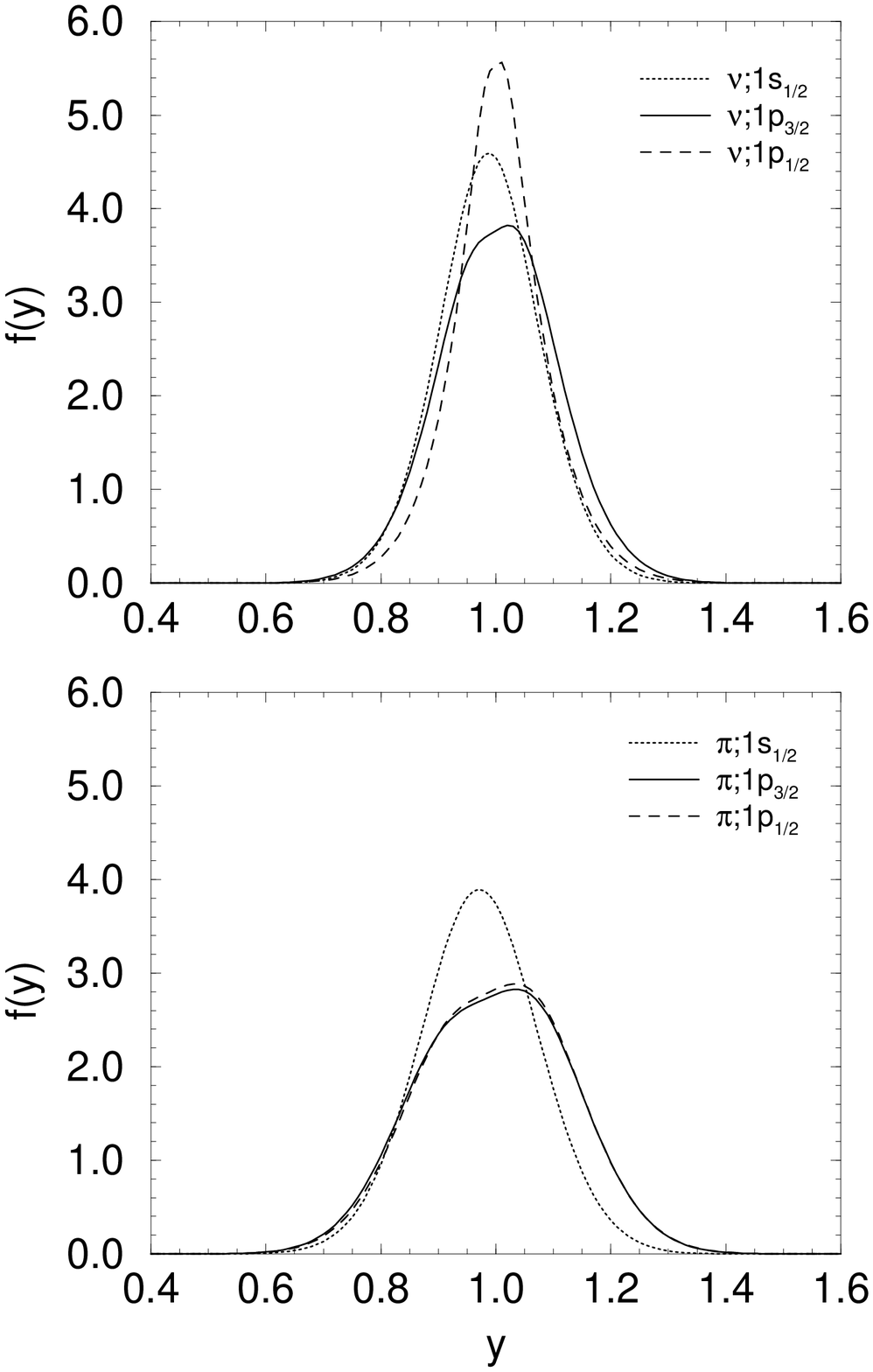,height=10cm}
\caption{Same as Fig.~\protect\ref{fig:y67} but 
for $^{11}$Li. The results shown in the left 
and right panels are for the cases of $S_n = 301$ keV and
350 keV, respectively.} 
\label{fig:y11}
\end{center}
\end{figure}

The longitudinal momentum distribution, $f_{\alpha_i/A}(y)$, must 
satisfy the condition of baryon number conservation: 
\begin{equation}
\int_0^{A} dy \, f_{\alpha_i/A}(y) = 1 .
\label{cond2}
\end{equation}
We have confirmed that $f_{\alpha_i/A}(y)$ obtained numerically 
satisfy the condition given in Eq. (\ref{cond2}) with an accuracy of
better than 0.2 \%. 
In Figs.~\ref{fig:y67} $-$ \ref{fig:y11} we present the $f(y)$ 
distributions for $^{6-11}$Li. Remarkably, the halo neutrons in $^{11}$Li 
are distributed quite sharply and 
the shape of $f(y)$ is almost symmetric around $y = 1$.  
This means that to a good approximation such neutrons can be regarded as 
being essentially free. 

\subsection{Numerical results for $F_2^A$}
\label{subsec:resultF2}

\begin{figure}
\begin{center}
\epsfig{file=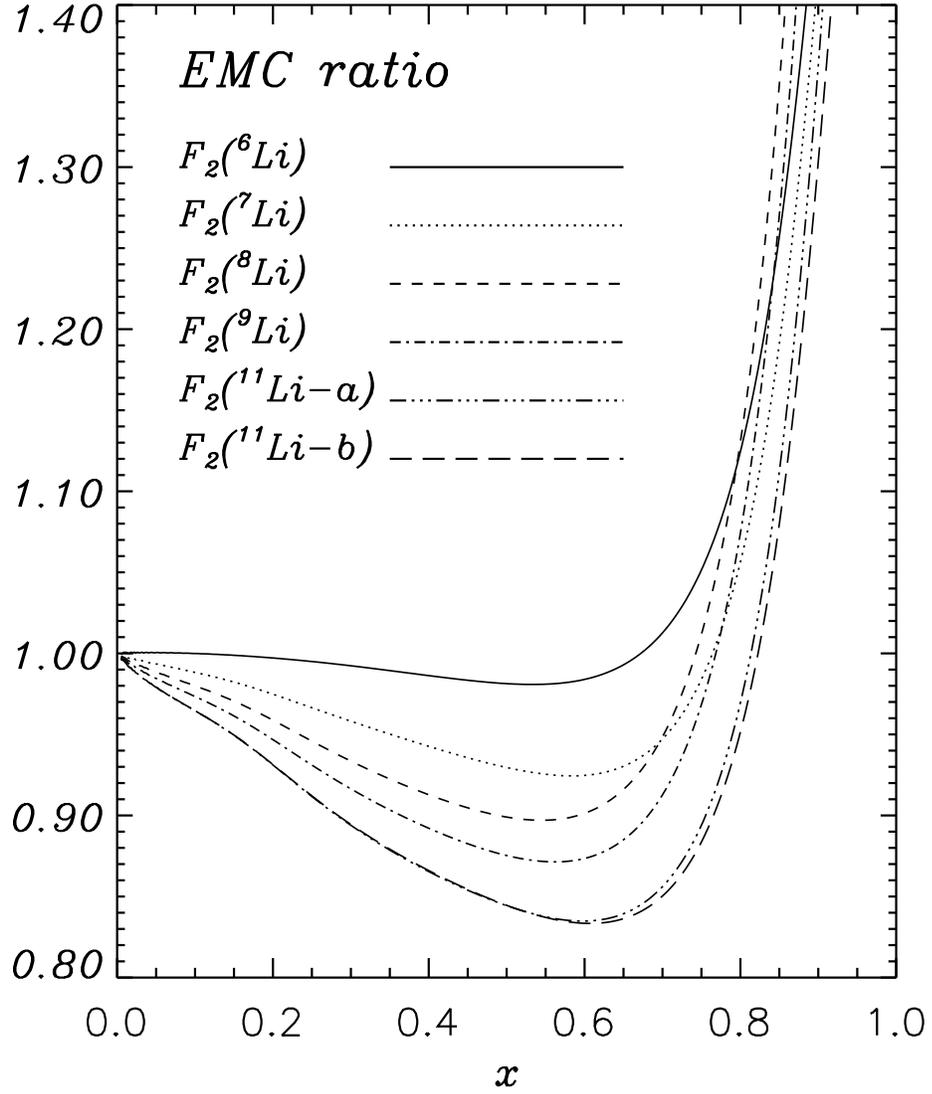,height=16cm}
\caption{Ratios of the structure functions of Li isotopes to 
that of the deuteron at $Q^2 = 10$ GeV$^2$ without isoscalarity 
corrections. $^{11}$Li-a(b) corresponds 
to the case of $S_n = 301 (350)$ keV. }
\label{fig:emc}
\end{center}
\end{figure}

Here we present numerical results for the spin-average structure 
functions of the Li isotopes, which were calculated using Eqs. (\ref{conv}) 
and (\ref{fin2fy}). For the structure function of the free nucleon,
the CTEQ5 parameterization~\cite{cteq5} was used. 
Fig.~\ref{fig:emc} illustrates the EMC ratios of the Li isotopes, which
are defined by $(F^{A}_2(x)/A)/(F^{d}_2(x)/2)$. 
We should note that the nuclear structure functions 
in the EMC ratio are usually corrected to be iso-scalar 
-- the so-called iso-scalarity correction~\cite{review,pw}. 
However, we did not apply the correction to the present ratio, and hence 
the Li structure functions in the numerator 
are not iso-scalar except for $^6$Li. 
The structure function of the deuteron
was calculated using the Paris potential~\cite{bissey}. 
In order to evaluate its recoil energy, $\langle T_R \rangle$,
we employed relativistic kinematics, since the residual debris is 
just one nucleon.  We find that 
$\langle T_R \rangle \simeq 9.0$ MeV and 
$-\varepsilon + \langle T_R \rangle \simeq$ 11 MeV~\cite{bissey1}. 
(If one chooses nonrelativistic kinematics $\langle T_R \rangle$ 
is about $9.4$ MeV~\cite{bissey1}.) 

In Fig.~\ref{fig:emc} each EMC ratio increases rapidly with $x$ 
in the region $x \geq 0.7$ and has a minimum at around $x = 0.6$. 
Also the minimum grows deeper as the number of neutrons increases.
The rapid increase of the EMC ratio is caused by the Fermi motion,
which can enable the nuclear structure function to have a tiny but finite
component at $x > 1$. 
Concerning the depression from unity appearing at intermediate $x$, 
one needs a careful discussion because no iso-scalarity
correction is applied in the present EMC ratio.

{}For $^6$Li, the depression from unity 
around $x=0.6$ is rather small. From Tables~\ref{tab:energy} 
and \ref{tab:kien} one can calculate the average-separation energies of 
protons, neutrons and all the nucleons, as well as the average-recoil energy 
for each Li isotope. They are presented in Table~\ref{tab:average}. 
The average proton- and neutron-separation energies,
$\langle \varepsilon \rangle_{p}$ and $\langle \varepsilon \rangle_{n}$, 
of $^6$Li are about 15 MeV and 17 MeV, respectively, while the 
average-recoil energy $\langle T_R \rangle$ is about 2.3 MeV. 
The nuclear binding effect, which is given by the sum of them, 
is thus about 18 MeV for $^6$Li, while it is about 11 MeV
for the deuteron. The small difference 
between these values is the 
reason for the small depression in the EMC ratio of $^6$Li.

{}For $\beta$-unstable Li isotopes, which are not iso-scalar,
the minima appearing around $x = 0.6$ in the EMC ratios become deeper 
as the neutron number increases. Let us consider the case of $^{11}$Li 
as an example.  From Table~\ref{tab:average} 
we find $\langle \varepsilon \rangle_n$ = 6.0 (5.9) MeV, 
$\langle \varepsilon \rangle_p$ = 30 (28) MeV, the average-separation 
energy of the total system, 
$\langle \varepsilon \rangle_{tot} 
\equiv \frac{1}{A} 
(N \langle \varepsilon \rangle_n + Z \langle \varepsilon \rangle_p$) 
= 12 (12) MeV, and $\langle T_R \rangle$ = 1.4 (1.3) MeV for $^{11}$Li-a(b). 
Thus, the nuclear binding of $^{11}$Li
is about 13 MeV, which is much closer to that of the deuteron, 
compared with the case of $^6$Li. It is then expected that 
the depression in the EMC ratio of $^{11}$Li would be more suppressed 
than that of $^6$Li. 
However, the EMC ratio shown in Fig.~\ref{fig:emc} is completely opposite. 
Since $\beta$-unstable Li isotopes have more neutrons than protons, 
as the neutron number increases the neutron structure function 
should contribute more to the EMC ratio than does the proton. 

Let us roughly estimate the effect of neutron excess on the EMC 
ratio.  We suppose that a nucleus is a simple collection of protons 
and neutrons and that the binding effect is ignored. It is well known that 
$F_2^p(x) > F_2^n(x)$ and 
around $x = 0.6$ the structure function of the free neutron is about a half 
of that of the free proton~\cite{halzen}.  
In this case the EMC ratio of $^6$Li 
to the deuteron is just unity because $^6$Li is iso-scalar.  
Using the relationship $F_2^p(x) \simeq 2 F_2^n(x)$ 
one finds that the ratios for $^{7,8,9,11}$Li become 0.95, 0.92, 0.89 
and 0.85 at $x \simeq 0.6$, respectively. It is remarkable that 
those values are already close to the calculated ratios at $x = 0.6$ in 
the figure. When the effects of the nuclear binding and Fermi motion are
turned on, they, of course, suppress the Li structure function 
at intermediate $x$ and enhance it at large $x$. 
However, such a suppression at intermediate $x$ may not be large because 
it is approximately determined by the difference between 
the sum of the average separation and recoil energies of Li 
and that of the deuteron -- see Table~\ref{tab:average}. 

If those successive EMC ratios could be measured in the future 
one would obtain systematic information on 
the off-mass shell effect on the neutron 
structure function in a nuclear medium~\cite{off}. 
We also note that the EMC ratio 
of $^{11}$Li is not sensitive to the choice of the one-neutron separation 
energy, $S_n$ (see Fig.~\ref{fig:emc}). 

\begin{figure}
\begin{center}
\epsfig{file=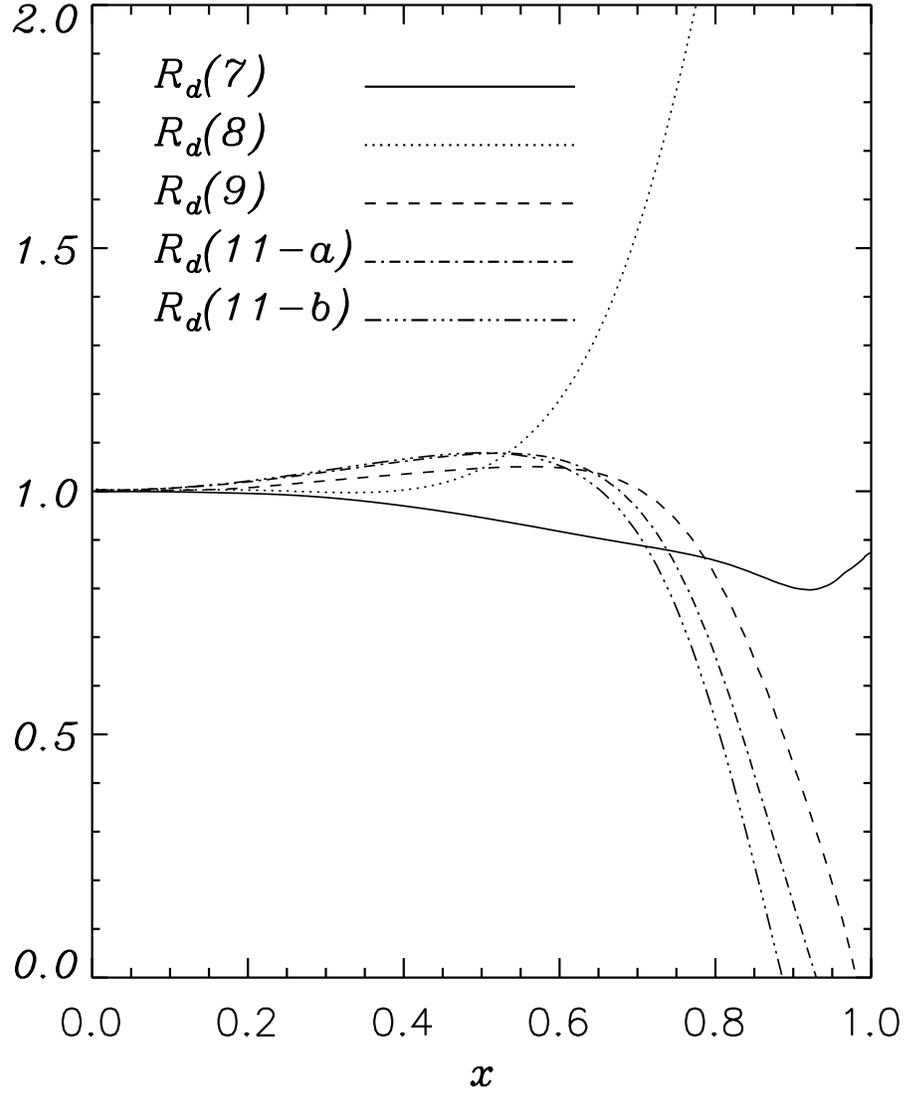,height=16cm}
\caption{$R_d(A)$ for Li isotopes ($Q^2 = 10$ GeV$^2$).  $R_d(11-a(b))$ 
is for the case of $^{11}$Li with $S_n = 301 (350)$ keV. }
\label{fig:lideut}
\end{center}
\end{figure}

We next discuss how the last neutron in each Li isotope is {\em bound}
as compared with the neutron in the deuteron using the following
ratio~\cite{sai1}
\begin{equation}
R_d(A) = \frac{F_2(^A{\rm Li}) - F_2(^{A-1}{\rm Li})}{F_2^d - F_2^p} , 
\label{Rd}
\end{equation}
where $F_2(^A{\rm Li})$ is the nuclear structure function of $^A$Li.  
Note that for $R_d(11)$ take the numerator to be 
$[F_2(^{11}{\rm Li}) - F_2(^{9}{\rm Li})]/2$. 
The results are plotted in Fig.~\ref{fig:lideut}.
In the region $x < 0.7$ the ratios $R_d(A)$ exceed unity, except for $A=7$.
This behavior may reflect the tendency for the energy level of 
the neutron-$1p_{3/2}$ orbit to decrease with increasing neutron number
(see Table~\ref{tab:energy}), even though the odd-even mass number effect 
can be seen in those energy levels.
In large $x$ region $R_d(A)$ becomes less than unity, which is 
consistent with the fact that the momentum distribution of the last neutron
in the Li isotope does not have a large high-momentum component. 
However, one has to keep in mind that 
the contribution of the nucleons in the core nucleus to 
the structure function $F_2(^A{\rm Li})$ is not completely cancelled 
by $F_2(^{A-1}{\rm Li})$ and that there remains an unavoidable 
contamination of the binding effect on the proton in the deuteron 
in the denominator.

\begin{figure}
\begin{center}
\epsfig{file=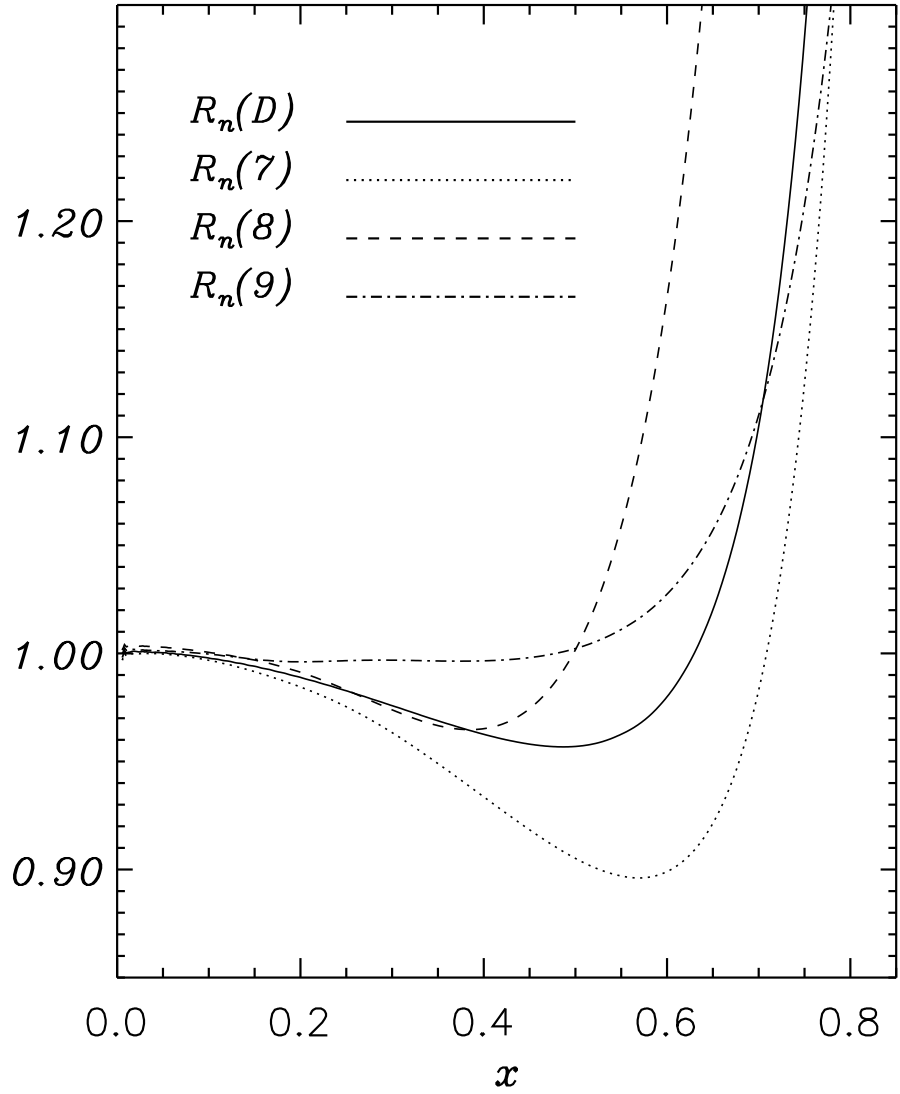,height=16cm}
\caption{$R_n(A)$ for $^{6-9}$Li isotopes ($Q^2 = 10$ GeV$^2$). The ratio of 
the difference between $F_2^d$ and $F_2^p$ to $F_2^n$ is also presented 
(denoted by $R_n(D)$). }
\label{fig:lin69}
\end{center}
\end{figure}

\begin{figure}
\begin{center}
\epsfig{file=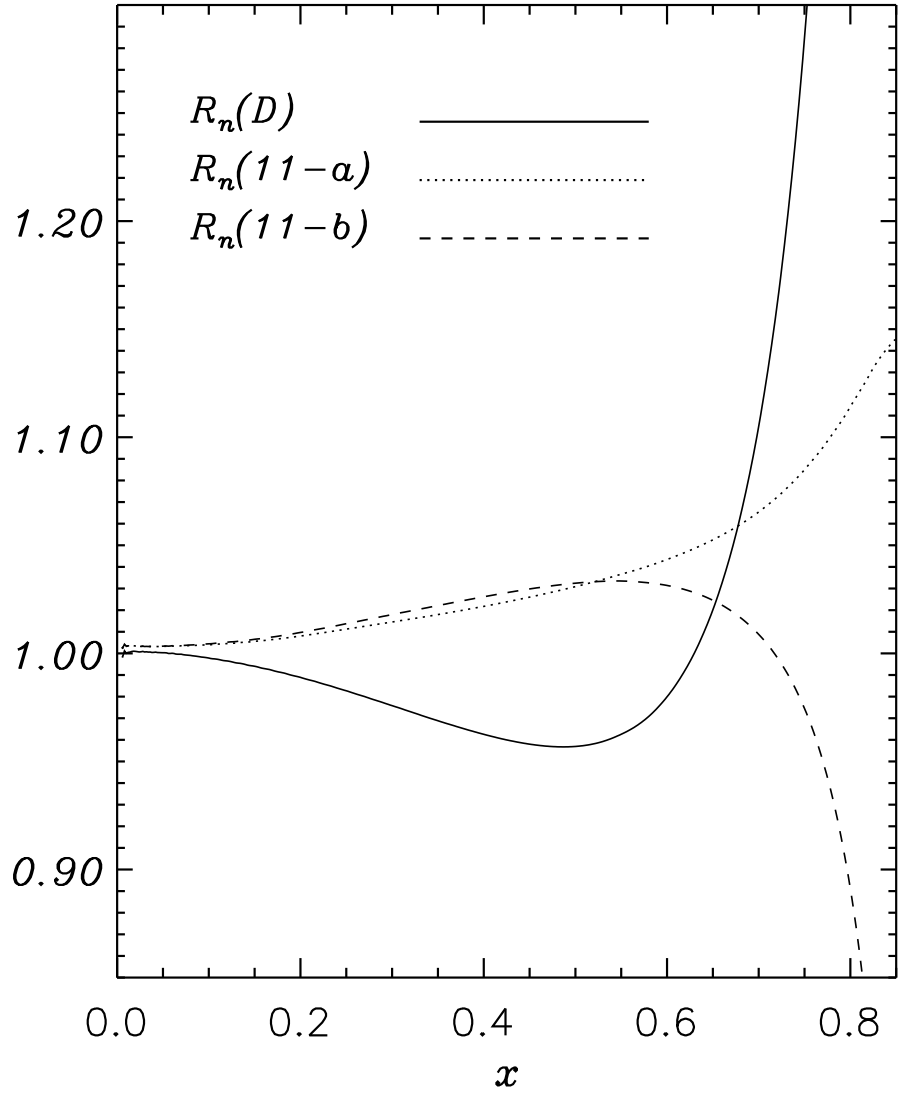,height=16cm}
\caption{$R_n(A)$ for $^{11}$Li ($Q^2 = 10$ GeV$^2$).  $R_n(11-a(b))$
is for the case of $^{11}$Li with $S_n = 301 (350)$ keV. The ratio 
$R_n(D)$ is also presented for comparison. }
\label{fig:lin11}
\end{center}
\end{figure}

Let us turn to the discussion of the following ratio~\cite{sai1}
\begin{equation}
R_n(A) = \frac{F_2(^A{\rm Li}) - F_2(^{A-1}{\rm Li})}{F_2^n} . 
\label{Rn}
\end{equation}
It would be useful to extract the neutron structure function from data 
for the series of Li isotopes. (For $A = 11$ we set the same numerator  
as in $R_d(11)$.) 
If one uses the ratio, $R_n(A)$, and the observed structure functions of 
$^A$Li and $^{A-1}$Li, then one can obtain the neutron structure function 
from $[F_2(^A{\rm Li}) - F_2(^{A-1}{\rm Li})]_{obs.} / R_n(A)$.  
Figs.~\ref{fig:lin69} and \ref{fig:lin11} 
show $R_n(A)$ for Li isotopes as well as that 
for the deuteron, $R_n(D) = (F_2^d - F_2^p)/F_2^n$.  
In the case that the neutron structure function is assumed to be simply given 
by $F_2^d - F_2^p$, one can see from $R_n(D)$ that the correction 
due to the nuclear binding is about $4$ \% at around $x = 0.5$ and 
that the effect of Fermi motion
is quite large in the region $x > 0.7$ (about 10 \% at $x = 0.7$). 

On the other hand, some combinations of the Li structure functions 
provide a pretty flat ratio below $x \sim 0.6 - 0.7$. 
Within the present model, for example, the ratio $R_n(9)$ is almost unity 
for $x < 0.5$ and exceeds unity 
by only about 2 \% at $x = 0.6$. We find in Table~\ref{tab:average} 
that the average-separation energies 
as well as the average-recoil energies of $^8$Li are almost identical to 
those of $^9$Li.  This is the reason why $R_n(9)$ is so close to unity 
at small and intermediate $x$.  
The ratios $R_n(11-a)$ and $R_n(11-b)$ are also 
flat below $x < 0.5 - 0.6$ because of the weak binding of the 
neutrons in the halo of 
$^{11}$Li\footnote{Recall that the cases ``$^{11}$Li(a-b)" correspond to the 
two choices of separation energy introduced earlier, 
$S_n = 301$ (a) or $S_n = 350$ (b) keV.}
(See also Table~\ref{tab:param}.) The ratios are about $1.03$ 
at $x = 0.5$.  In particular, in $R_n(11-b)$
the deviation from unity, $|R_n(11-b) -1.0|$, is below  
$3$ \% for $x < 0.75$.  

However, in the large $x$ region the ratio increases in $R_n(11-a)$, 
while it decreases in $R_n(11-b)$.  For the case $S_n = 301$ keV 
($^{11}$Li-a) 
the valence neutrons in $^{11}$Li are bound more loosely than 
those for the case of $S_n = 350$ keV ($^{11}$Li-b). Correspondingly, 
the rms radii of the neutrons in the $1p_{1/2}$-orbit are, 
respectively, 4.91 fm 
and 4.82 fm for the former and latter cases 
(see Table~\ref{tab:energy}).  In order to reproduce the measured rms 
radius of $r_{^{11}{\rm Li}} = 3.10$ fm, 
the nucleons in the core need to be
more tightly bound for the former case than for the latter  
(see again Table~\ref{tab:energy}). Thus, the effect of Fermi 
motion due to the nucleons in the core appears to be stronger for 
the former case than that for the latter, 
which makes $R_n(11-a)$ increase and $R_n(11-b)$ 
decrease in the region $x > 0.7$. From the behavior of $R_n(11)$ at large 
$x$, we can say that the ratio is quite 
sensitive to how loosely the neutrons in the halo of $^{11}$Li are bound, 
and that the ratio strongly depends 
on high momentum components in the nucleon momentum distributions for $^9$Li 
and $^{11}$Li. It is certainly intriguing to 
work out a more elaborate calculation for the nuclear structure of the Li 
isotopes, where residual interactions such as pairing 
correlations are taken into account, in order to study 
the ratio $R_n(A)$ in detail. 

\begin{figure}
\begin{center}
\epsfig{file=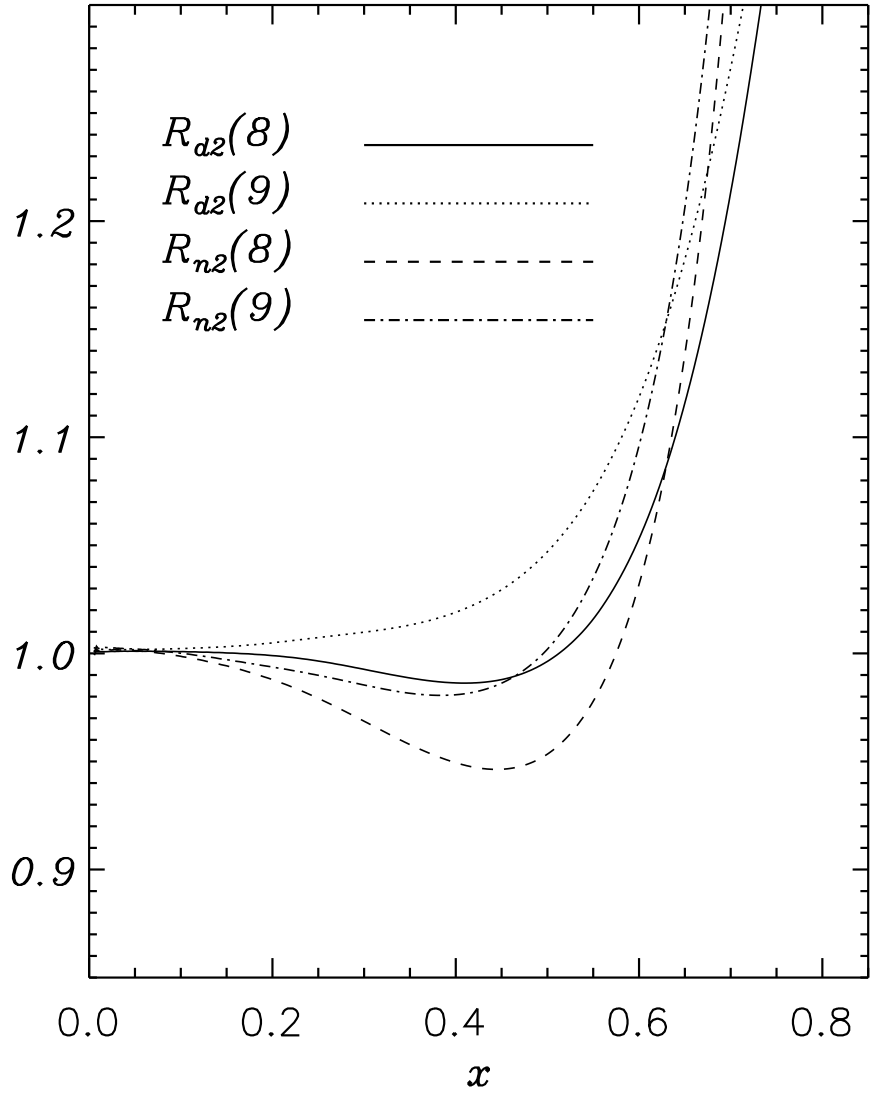,height=16cm}
\caption{$R_{d2}(A)$ and $R_{n2}(A)$ for $^{8,9}$Li ($Q^2 = 10$ GeV$^2$).}
\label{fig:com}
\end{center}
\end{figure}

We have already noticed that the energy spectra of the Li isotopes show 
the odd-even mass number effect (see Table~\ref{tab:energy}). 
As a method for cancelling the effect simply, it would be interesting to see 
the following ratios: 
\begin{eqnarray}
R_{d2}(A) &=& \frac{F_2(^A{\rm Li}) - F_2(^{A-2}{\rm Li})}
{2(F_2^d - F_2^p)} ,
\label{Rd2} \\
R_{n2}(A) &=& \frac{F_2(^A{\rm Li}) - F_2(^{A-2}{\rm Li})}{2F_2^n} . 
\label{Rn2} 
\end{eqnarray}
Fig.~\ref{fig:com} presents the ratios for $^8$Li and $^9$Li.
(Note that $R_{d2}(A=11) = R_{d}(11)$ and $R_{n2}(11) = R_{n}(11)$.) 
One can again see that the ratios for $x < 0.6$ are pretty close to unity 
and the rapid increase due to the Fermi motion appears for 
$x > 0.6$. In particular, the deviation of $R_{d2}(8)$ ($R_{n2}(9)$) 
from unity lies within 2\% (3\%) for $x < 0.55$.  
Interestingly, $R_d(9)$ decreases in the large $x$ region, 
while $R_{d2}(9)$ increases there. It means that the $^8$Li nucleus 
has relatively larger high-momentum components than $^7$Li,
which may partly result from the fact that the binding energy of 
protons in the $1s_{1/2}$ orbit for $^8$Li is much larger than 
that for $^7$Li (see Table~\ref{tab:energy}). 

\section{Spin structure functions of Li isotopes}
\label{sec:spin}

Contrary to the case of unpolarized structure functions~\cite{review,pw},
so far, only a few theoretical studies have been made for the effect of 
nuclear binding on the spin-dependent structure
functions~\cite{off,bissey,Guzey,Close,Indumathi,ciofi,deutspin,kp}. 
The present situation seems to be that the errors in 
extracting nuclear effects from the 
spin-dependent structure functions still seem to be within
the experimental error bars. However, recent progress in 
high precision measurements of spin-dependent structure functions for 
the neutron, proton, and deuteron~\cite{recentspin,g1exp}, 
together with improved parameterizations for the polarized parton 
distribution functions (PPDF)~\cite{FS,GRSV}, will make it necessary 
to know the effect with a high accuracy in near future. 

\subsection{Nuclear spin structure functions}
\label{subsec:spinfunc}

Here we systematically consider the spin 
structure functions of the Li isotopes. 
Let us suppose that the spin-dependent structure functions of a nucleus 
separate into the contributions from those of protons and 
neutrons. In the conventional nuclear model the polarization of 
a nuclear ground state 
is determined by the unpaired nucleons because two identical 
nucleons in the same shell orbit prefer coupling to spin of zero.
Thus, the spin of a Li isotope with an odd mass number is $3/2$, 
which is attributed only to the valence proton in the $1p_{3/2}$ orbit, 
while the spin of the $^{6 (8)}$ Li ground state is 1 (2), which 
is given by a combination of the unpaired proton and neutron in the
$1p_{3/2}$ orbit (see Table~\ref{tab:exp}).

It is, however, important to consider residual interactions between
single-particle states in a nucleus, 
which lead the configuration mixing 
and hence a change of the wave function of the valence nucleon.  If 
we ignore the nuclear binding and Fermi motion effects 
the nuclear spin structure function may be approximately given 
by~\cite{Guzey,ciofi} 
\begin{equation}
g_1^A(x) = P_n^A \times g_1^n(x) + P_p^A \times 
g_1^p(x) , 
\label{g1A}
\end{equation}
where $P_{p (n)}^A$ is the effective polarization of proton (neutron) in 
the nucleus $A$. It is well known that, for example, in $^3$He the 
tensor force etc. provide a mixture of $S'$- and $D$-states 
to the $S$-wave state and hence that the neutron is somewhat depolarized and 
the proton gains a slight downward (negative) polarization due to the 
configuration mixing: $P_n^{A=3} = 86 \pm 2 \%$ and 
$P_p^{A=3} = 2 \times (-2.8 \pm 0.4) \%$~\cite{ciofi}.  In the series of Li 
isotopes a similar situation would occur because of a mixing of higher 
configurations. 
It is therefore necessary to perform a calculation of the nuclear 
structure of Li isotope including residual interactions to get 
precise values of the effective polarizations. 
We do not perform such a calculation in this paper. 

Instead, in this section we consider the effects of nuclear binding and 
Fermi motion on the spin asymmetry for
a target with spin $3/2$ in detail. This corresponds to DIS 
off $^{7,9,11}$Li targets. For such targets there are 
four non-vanishing multipole structure functions in the scaling limit: 
$^{3/2}_{~0}F_1$, $^{3/2}_{~2}F_1$, $^{3/2}_{~1}g_1$ and 
$^{3/2}_{~3}g_1$ -- recall the discussion in Sec.~\ref{sec:intro}. 
The first one is proportional to the usual spin-average 
structure functions, which has been studied in the previous 
section, while the second one is analogous to the quadrupole 
structure functions, $b_1$, for a spin-1 target. 
The last two provide information on quark spin 
asymmetries: $^{3/2}_{~1}g_1$ is the analog to $g_1$ for a spin-1/2 target, 
while $^{3/2}_{~3}g_1$ is a new asymmetry function. 
It is important to keep in mind that in the present approximation only 
unpaired valence nucleons contribute to the multipole spin structure 
functions. 

\subsection{Spin structure function for a spin-3/2 target}
\label{subsec:spin32}

The spin asymmetry for a spin-$3/2$ Li target is given by~\cite{jaffe} 
\begin{equation}
\frac{d \Delta \sigma^{3/2}}{dxdy} = C x [1 - (1-y)^2]
       \sum_{L=1,3} {^{3/2}_{~L}g_1(x)} \rho_L^{3/2} , \label{spinasym}
\end{equation}
where the factor $C$ is given below Eq. (\ref{cs2}). 
Then, we can calculate the multipole spin structure function using  
the following convolution form: 
\begin{equation}
^{3/2}_{~L}g_1(x) = \int_x^A dy \, \frac{1}{y} \, g_L^{3/2}(y) 
g_1^p(x/y) , \label{convJ}
\end{equation}
where $g_L^{3/2}(y)$ is the light-cone spin distribution given by 
${\rm tr} [M_L^{3/2} g_1^{3/2}]$ -- note that 
$g_1^{3/2}$ is a diagonal $(2J+1) \times (2J+1)$ matrix defined below 
Eq. (\ref{multifg}). 

We find $g_L^{3/2}$ explicitly as 
\begin{eqnarray}
g_1^{3/2}(y) &=& \frac{2\sqrt{5}}{3} c_0(y) - \frac{4}{3\sqrt{5}} c_2(y), 
\label{gc1} \\
g_3^{3/2}(y) &=& -\frac{6}{\sqrt{5}} c_2(y), \label{gc3}
\end{eqnarray}
where 
\begin{eqnarray}
c_L(y) &=& \int \frac{d^4k}{(2\pi)^4} 2\pi \delta(k_0 - {\bar M}) 
| \Psi_{1p_{3/2}}^p({\vec k})|^2 \left[ 1 + \frac{k^3}{k^0} 
- \frac{{\vec k}\,^2 - (k^3)^2}{k^0(k^0 + M)} \right] \nonumber \\
&\times& P_L(\cos\theta) 
\delta \left( y - \frac{k^0 + k^3}{M} \right) . \label{cL}
\end{eqnarray}
In Eq. (\ref{cL}) the term of the square blacket is the so-called 
flux factor, which stems from the matrix element of $\gamma^+\gamma_5$ 
with respect to the nucleon spinor. In the limit $k^0 \to M$ 
it coinsides with the flux factor given in Refs.~\cite{pw,kp}. 
Instead of the present flux factor, the factor, $(1 + k^3/k^0)$, which 
is the same as that for the $f(y)$ distribution, is often used to 
calculate the nuclear spin structure function. 
However, the difference between two cases is less than
2\% in the $y$ distribution (see Fig.~\ref{fig:spin1} in 
subsection~\ref{subsec:resultg1}).

After performing the integrals with respect to $k^0$ and the angular 
variables we finally obtain 
\begin{eqnarray}
c_0(y) &=& \frac{M^2}{{\bar M}} \int_{k_L}^\infty \frac{k dk}{2(2\pi)^3} 
\vert \phi_{1p_{3/2}}^p(k) \vert^2
\left[ y + \frac{1}{2} ( y - 1)^2 - \frac{k^2}{2M^2} \right] , 
\label{c0} \\
c_2(y) &=& \frac{M^2}{{\bar M}} \int_{k_L}^\infty \frac{k dk}{4(2\pi)^3}
\vert \phi_{1p_{3/2}}^p(k) \vert^2 
\left[ 3 \frac{M^2}{k^2} \left( y - \frac{{\bar M}}{M} \right)^2 - 1 \right] 
\left[ y + \frac{1}{2} ( y - 1)^2 - \frac{k^2}{2M^2} \right] , 
\label{c2}
\end{eqnarray}
where $\phi_{1p_{3/2}}^p(k)$ is the radial wave function of the 
valence proton in momentum space.  The term of 
${\cal O}(\varepsilon/M)$ is neglected in the flux factor because its 
contribution is very small. 

\subsection{Numerical results for spin structure functions}
\label{subsec:resultg1}

\begin{figure}
\begin{center}
\epsfig{file=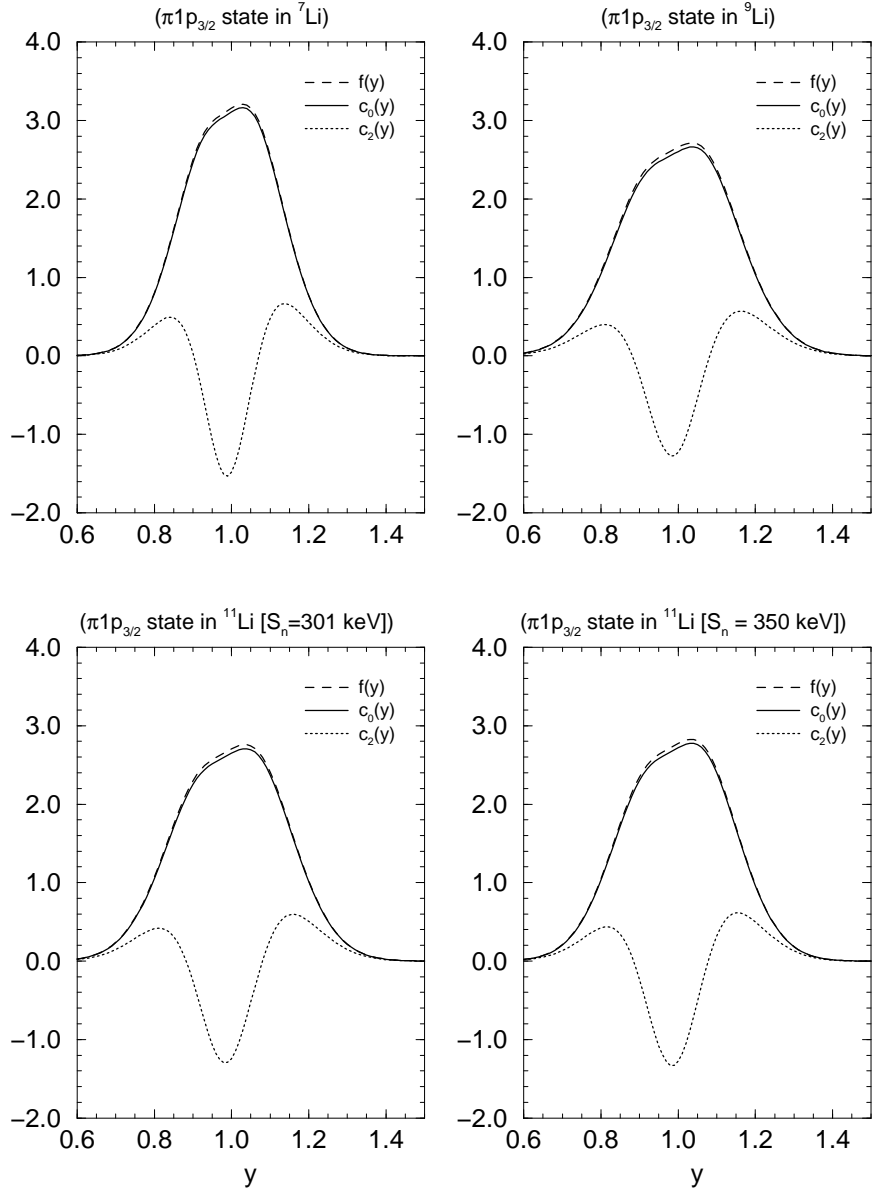,height=16cm}
\caption{Distributions, $c_0(y)$ and $c_2(y)$, for 
$^{7,9,11}$Li. For comparison, $f(y)$ is also presented. 
}
\label{fig:spin1}
\end{center}
\end{figure}

In Fig.~\ref{fig:spin1} we plot the distributions, 
$c_0(y)$ and $c_2(y)$, for $^{7,9,11}$Li isotopes. For comparison, 
we also present the light-cone momentum distribution, $f(y)$, of 
the valence proton in the $1p_{3/2}$ orbit. One sees that 
the shape of the $c_0(y)$ distribution is very similar to the
$f(y)$ distribution, while that of $c_2(y)$ is quite different. 
Note that the small difference between $c_0(y)$ and $f(y)$ is caused by 
the second and third terms of the flux factor in Eq. (\ref{c0}). 
It should be noted that $c_L(y)$ for $^{11}$Li is not sensitive to
the choice of the one-neutron separation energy, $S_n$, and hence that
the difference between the $c_L(y)$ distributions for $S_n = 301$
and $350$ keV is quite small. 

There is a sum rule for the multipole spin structure 
function~\cite{jaffe}: 
\begin{equation}
\int_0^A dx \, x^{n-1} \, \, {^J_Lg_1(x)} = 0 , \ \ \ \mbox{for} \ \ 
1 \le n < L . \label{sumg}
\end{equation}
This offers a sum rule for $c_2(y)$  
\begin{equation}
\int_0^A dy \, c_2(y) = 0 ,  \label{sumc}
\end{equation}
while $c_0(y)$ is free from such a constraint.  We have 
checked the sum rule for $^{3/2}_{~3}g_1 (n=1)$ numerically, and found that 
the absolute value of the integral, Eq. (\ref{sumg}), is less than $0.0014$ 
for the $^{7,9,11}$Li isotopes, 
where the calculation was performed in the region 
$10^{-3} \le x \le 2.20$.  The sum rule is certainly satisfied 
within the numerical precision.

\begin{figure}
\begin{center}
\epsfig{file=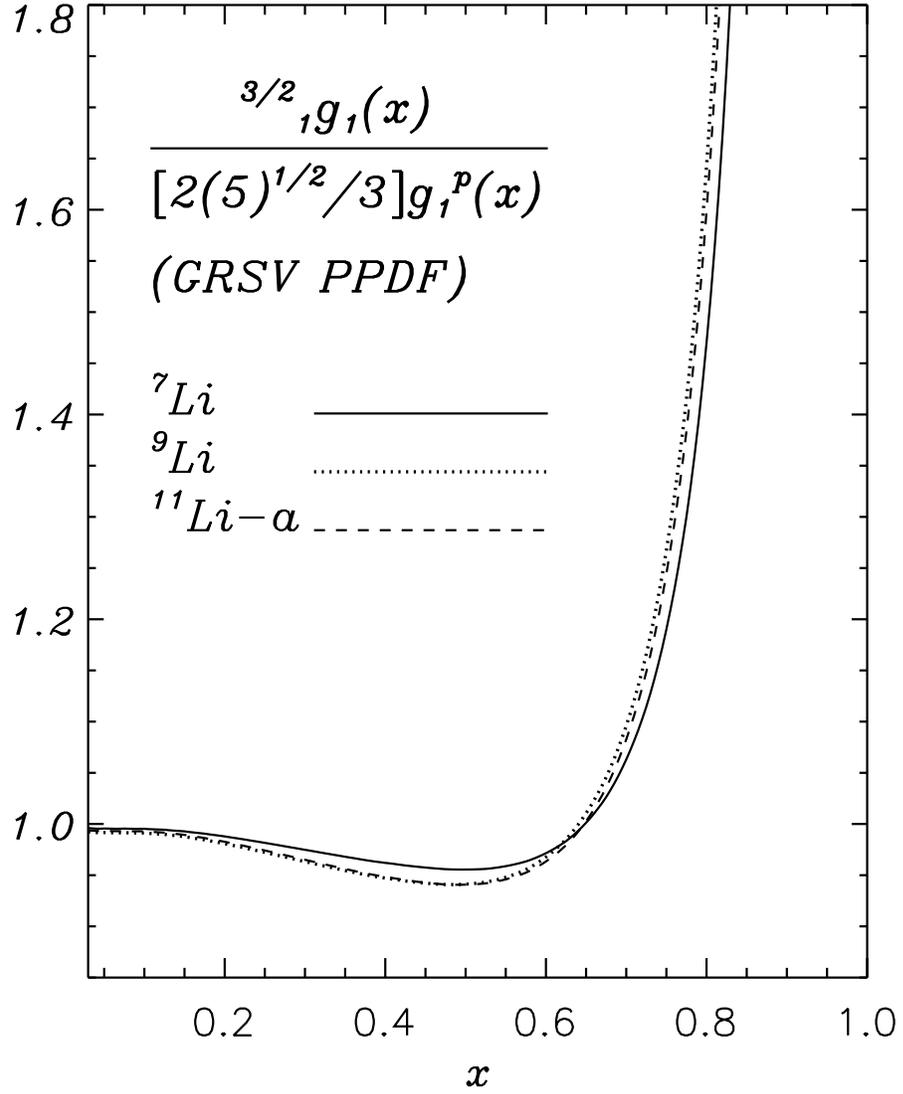,height=16cm}
\caption{Ratio of the multipole spin structure function, $^{3/2}_{~1}g_1$, to 
the usual proton spin structure function, $g_1^p$.  The geometrical factor 
of $2\sqrt{5}/3$ is included in the denominator of the ratio. }
\label{fig:gLi1}
\end{center}
\end{figure}

\begin{figure}
\begin{center}
\epsfig{file=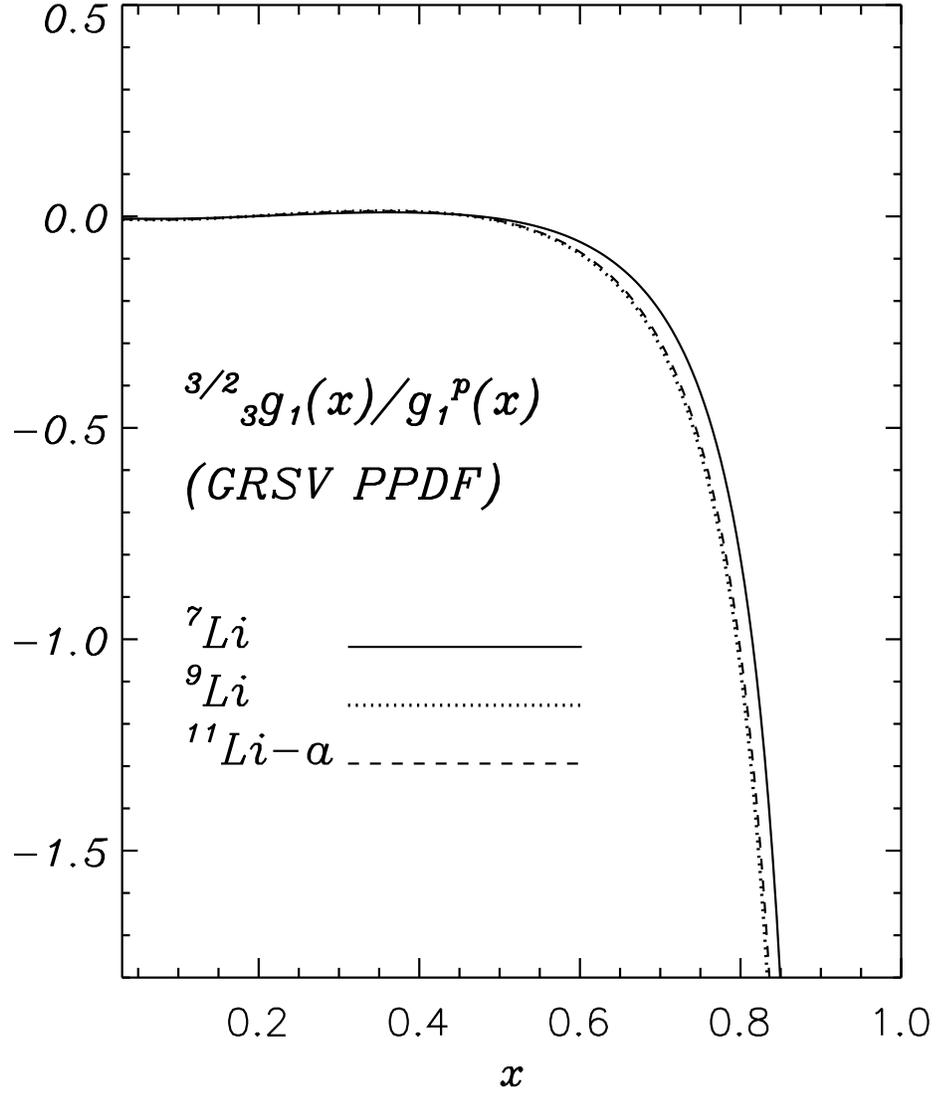,height=16cm}
\caption{Ratio of $^{3/2}_{~3}g_1$ to $g_1^p$.}
\label{fig:gLi3}
\end{center}
\end{figure}

In Figs.~\ref{fig:gLi1} and \ref{fig:gLi3} we present the multipole 
spin structure functions, $^{3/2}_{~1}g_1$ and $^{3/2}_{~3}g_1$, 
for $A= 7,9,11$ 
normalized $g_1^p$ for the free proton. For the numerical 
calculation we have used the polarized parton distribution functions (PPDF) 
parameterized by D. de Florian {\it et al.}~\cite{FS} and 
M. Gl\"{u}ck {\it et al.}~\cite{GRSV} (denoted by GRSV PPDF).
We use one of the options for both, up to the next-to-leading-order (NLO)    
within the $\overline{\rm MS}$ scheme, which includes the effects of flavor 
asymmetries in the proton sea. The calculations are performed at 
$Q^2 = 10$ GeV$^2$. Since the two parameterizations eventually gave very 
similar results we show only the result with the GRSV parameterization in 
the figures.  Moreover, since the choice of the one-neutron separation 
energy, $S_n$, in $^{11}$Li does not affect the calculated result much, 
we illustrate only the ratios for $^{11}$Li-a (i.e., $S_n = 301$ keV). 

In graphing the ratio for $^{3/2}_{~1}g_1$ (see Fig.~\ref{fig:gLi1}) we have 
divided out the ``geometrical factor'' $2\sqrt{5}/3$ so the plotted ratio 
would be unity in the absence of the nuclear binding and Fermi motion. 
The shape of the ratio is very similar to the usual EMC ratio -- see 
also the distributions of $f(y)$ and $c_0(y)$ in Fig.~\ref{fig:spin1}.  
The effect of the nuclear binding and Fermi motion on $^{3/2}_{~1}g_1$ is then 
about $10 \%$ for $x < 0.7$, while the Fermi motion effect is very 
important for large $x$.  This observation is consistent with the 
earlier results for the deuteron~\cite{deutspin,kp} and 
$^3$He~\cite{ciofi}. For $x < 0.6$ the ratio for $^7$Li is a little 
shallower than that for $^{9,11}$Li. This is because the weak binding 
of the valence proton in $^7$Li as compared with in $^{9,11}$Li isotopes 
-- see Table~\ref{tab:fit}. 

The shape of the ratio for $^{3/2}_{~3}g_1$ is quite different from that for 
$^{3/2}_{~1}g_1$. It is negative for small $x$ but
it becomes positive for $0.2 < x < 0.5$.  However, the magnitude is 
very small in the region of $x < 0.6$.  For very large $x$ it is
again negative and its absolute value becomes large because of the Fermi
motion. Because of the weaker binding of the valence proton in $^7$Li than 
in $^{9,11}$Li, the magnitude of the ratio for $^7$Li for $x > 0.5$ 
is smaller than that for $^{9,11}$Li. 
We should note here that the present result is quite different from that of 
Jaffe and Manohar~\cite{jaffe}. In particular, the ratio for large $x$ 
is negative in our realistic calculation, while it is large and positive in 
Ref.~\cite{jaffe}.  This discrepancy is caused by  
an approximation used in Ref.~\cite{jaffe}, 
where the distributions of $c_0(y)$ and $c_2(y)$ are 
expanded in terms of $\delta$-functions. 
This amounts to neglecting the positive wings of $c_2(y)$ in 
Fig.~\ref{fig:spin1}, which is numerically inaccurate.

\section{Summary and conclusion}
\label{sec:summary}

We have studied the spin-average and spin-dependent structure functions 
of the lithium isotopes.  
First, we have calculated the wave functions within Hartree 
approximation. In the present calculation we used a phenomenological 
nuclear potential of the Woods-Saxon type, in which parameters are 
fitted so as to reproduce the 
separation energies of protons and neutrons, the single-particle excitation 
energies and the rms radius for each Li isotope.  
In the calculation, for $^{11}$Li 
the separation energy for the last neutron in the halo 
was taken to be either $S_n = 301$ or $350$ keV 
in order to examine the $S_n$-dependence
of the structure function. 
Next, using the wave functions we have individually calculated 
the light-cone momentum distribution of the nucleon in each shell model orbit. 
We have found that the $f(y)$ distribution of the valence neutron in 
$^{11}$Li is very sharp and almost symmetric around $y = 1$, which means 
that this neutron is bound under a condition quite close to the 
neutron in free space. 

Adopting the usual convolution technique~\cite{review} we have calculated 
the spin-average structure functions of Li isotopes. The EMC ratios of the Li 
nuclei were first illustrated and the role of the 
neutron structure function in the EMC ratio was discussed.  As the neutron 
number increases the neutron structure function plays a more important role 
in the nuclear structure function because its weight becomes larger than 
that of the protons.  Furthermore, we have presented the ratio, $R_d$, which 
is defined by the ratio of 
the difference between the Li structure functions of mass number $A$ and 
$A-1$ to the difference between the structure functions of the deuteron 
and the free proton.  The ratio, $R_n$, which is defined as in $R_d$ but the 
denominator is the free neutron structure function, was also illustrated. 
The ratios show the sensitivity of 
the nuclear structure function to the 
last valence neutron. In particular, the ratios $R_n(9)$ and $R_n(11)$ 
are very useful for extracting the free neutron structure function from  
Li data.  We have also investigated the ratios $R_{d2}$ and 
$R_{n2}$, where the numerators are chosen to be $F_2(^{A}$Li$) - 
F_2(^{A-2}$Li$)$, to eliminate the odd-even mass number effect in the 
Li structure. 

We have studied the spin-dependent structure functions of Li 
isotopes.  In the present study we have concentrated on Li nuclei with 
odd $A$ and calculated the multipole spin structure functions, 
$^{3/2}_{~1}g_1$ and $^{3/2}_{~3}g_1$, where the former is the analog to 
$g_1^p$ for a spin-1/2 target and the latter is a new spin asymmetry which 
first arises for a spin-3/2 target.  We have studied the nuclear binding 
and Fermi motion effects on those spin structure functions in detail. 
Here the spin structure function is assumed to be given 
by a convolution form of the light-cone spin distribution of the 
valence proton, which is provided by a combination of $c_0(y)$ and $c_2(y)$, 
and the spin structure function of the free proton.  

We have found that the shape of the ratio of $^{3/2}_{~1}g_1$ to $g_1^p$ 
is very similar to the usual EMC ratio, while the ratio for $^{3/2}_{~3}g_1$ 
is quite different from the EMC one. It has been confirmed that the 
effect of the nuclear binding and Fermi motion on $^{3/2}_{~1}g_1$ is 
about $10 \%$ for $x < 0.7$, but that the Fermi motion effect is very
important for large $x$.  On the other hand, the multipole spin structure 
function of $^{3/2}_{~3}g_1$ is negative for small $x$ but
it turns to be positive for $0.2 < x < 0.5$.  For very large $x$ it is
again negative and the magnitude of the ratio 
becomes large because of the Fermi
motion. The present observation contradicts an earlier reported 
result, where an approximation was used to calculate the distributions 
of $c_0(y)$ and $c_2(y)$.  

We should note that in both the spin-average and spin-dependent cases 
a shadowing (and anti-shadowing) correction would be necessary for 
the structure functions at small $x (< 0.2)$.   
This would require delicated calculation along the 
line of Refs.~\cite{spin3,Guzey}. 
For further studies it will also be necessary to include two-body 
residual correlations in the calculation of the structure of Li. 

Here we would like to comment on the modification of the Gottfried
and Bjorken integrals in nuclei: the former is defined by an 
integral of the difference between the spin-average structure functions of 
a pair of mirror nuclei (divided by $x$)~\cite{sai1,vadim,sai2}, while 
the latter is given by an integral of the difference between
the spin structure functions of mirror nuclei~\cite{spin3,Guzey}.  
As mentioned in the 
introduction, such quantities could provide us significant information
on the {\em flavor-nonsinglet} parton distributions in a nucleus.
Together with the {\em flavor-singlet} structure functions, such 
information is quite important for determining the individual isospin
distributions in a nucleus separately.

We list several candidates for such pairs of mirror nuclei.
The most promising candidate is of course the pair $^3$He and
$^3$H~\cite{sai1,vadim,sai2}, where both the nuclei are 
stable under strong interactions. However, 
this is also the lightest pair, so that
the nuclear modification on the structure functions cannot be expected to be
large. The pair $^7$Li and $^7$Be should also be good because $^7$Be has a 
long half-life time (53 days) and $^7$Li is stable.  
We thus believe that the Li-Be pair may be the best one for testing 
the nuclear effects on the Gottfried and Bjorken integrals in a medium. 
The next candidates would be the pairs $^{11}$C--$^{11}$B and 
$^{13}$N--$^{13}$C, where the half-lives of $^{11}$C and $^{13}$N
are about 20 and 10 minute, respectively, while $^{11}$B and $^{13}$C are
stable.  For other candidates one can find some pairs, whose half-life
times are longer than 1 minutes: 
$^{14}$O (71 sec)--$^{14}$C (5730 years), 
$^{15}$O (120 sec)--$^{15}$N (stable),
and $^{17}$F (65 sec)--$^{17}$O (stable).
However, it is nearly impossible to measure
their structure functions with fixed-target experiments.

If DIS with high energy electrons (or muons) off unstable nuclei (like 
the series of Li isotopes) is realized by using a collider machine 
(like MUSES or (Tesla)HERA) in the future~\cite{muses,future,future2}, 
available mirror pairs would be widely extended and then
quark nuclear physics could be developed quite systematically.
Although we have discussed only the series of Li isotopes in this paper, 
we can point out other interesting radioactive isotopes: for example, 
the series of isotopes of Be, B, C, O, Na and Mg 
are of great promise~\cite{muses}. 
In particular, the series of oxygen isotopes are interesting. 
In the series the smallest and largest mass (neutron) numbers are 13 (5)
and 24 (16), respectively. For neutron-rich isotopes, a neutron skin or
a neutron halo is expected to be formed around the tightly bound
$^{16}$O core.
If we could control the atomic number and the difference between 
the proton and neutron numbers freely in measuring the nuclear 
structure functions, it would stimulate a great deal of works that 
should lead to interesting new quark 
physics in nuclei~\cite{sai1}. 

\vspace{1cm}
We would like to thank F. Bissey for providing us the light-cone 
momentum distributions of the  
deuteron. We also acknowledge numerous valuable discussions with 
V. Guzey on the polarized structure functions and small $x$ physics. 
This work was partly supported by the 
Australian Research Council. M.U. acknowledges support from
Funda{\c c}{\~a}o de Amparo {\' a} Pesquisa do Estado de S{\~a}o Paulo 
(FAPESP). K.T. would like to acknowledge the hospitality at KFA J\"{u}lich, 
where some of the numerical calculations were carried out. 
The present calculation for spin-dependent structure functions
was performed using the FORTRAN codes for
the polarized parton distribution functions by D. de Florian and
R. Sassot~\cite{FS}, and M. Gl\"{u}ck, E. Reya, M. Stratmann, and
W. Vogelesang~\cite{GRSV}. 



\newpage


\begin{table}
\caption{Experimental data for Li isotopes~\protect\cite{data,69li}. 
The one-neutron, $S_n(Z,N)$, one-proton, $S_p(Z,N)$, and two-neutron, 
$S_{2n}(Z,N)$, separation energies of the nucleus $^AZ \ (A=N+Z)$ are 
defined by $S_n(Z,N) = B(Z,N)-B(Z,N-1)$, $S_p(Z,N) = B(Z,N)-B(Z-1,N)$ 
and $S_{2n}(Z,N) = B(Z,N)-B(Z,N-2)$, respectively,
where $B(Z,N)$ is the total binding energy of the nucleus $^AZ$. 
The rms radius of $^A$Li is denoted by $r_A$. The numbers in the 
parentheses stand for the error bars of the observed data.}
\label{tab:exp}
\begin{center}
\begin{tabular}{clccccc} \hline \cr
   & $J^{\pi}; T$ & $B$ (MeV) & $S_n$ (MeV) & $S_{2n}$ (MeV)
& $S_p$ (MeV) & $r_A$ (fm) \\
\hline
$^6$Li & 1$^+; 0$     & 31.995   & 5.664 & & 4.589 & 2.32 \\
       &              & (0.0005) & (0.0505) & & (0.0505) & (0.03) \\
$^7$Li & 3/2$^-; 1/2$ & 39.245   & 7.250 & 12.91 & 9.998 & 2.33 \\
       &              & (0.0005)  & (0.0009) & (0.0505) & (0.0015) & (0.03) \\
$^8$Li & 2$^+; 1$     & 41.277   & 2.033 & 9.283 & 12.45 & 2.37 \\
       &              & (0.0005) & (0.0010) & (0.0010) & (0.0305) & (0.03) \\
$^9$Li & 3/2$^-; 3/2$ & 45.341   & 4.064 & 6.096 & 13.93 & 2.32 \\
       &              & (0.0019) & (0.0024) & (0.0024) & (0.0090) & (0.02) \\
$^{11}$Li & 3/2$^-; 5/2$ & 45.642   & 0.326 & 0.301 & 15.30 & 3.10 \\
          &              & (0.0271) & (0.0422) & (0.0290) & (0.0971) & 
(0.22)\\ \hline
\end{tabular}
\end{center}
\end{table}

\vspace{1cm}


\begin{table}
\caption{The first 1/2$^-$ levels of $^{7,9}$Li and $^{7,9}$Be
nuclei. The values presented are those of the first excited states 
(except for $^9$Be) and correspond to the single-particle 
excitations specified in the fouth column. 
The value for $^9$Be is for the third excited state.
The spins, parities, and energy levels of the first and second excited states 
of $^9$Be are $1/2^+$ (1.684 MeV) and $5/2^-$ (2.429 MeV), respectively.}
\label{tab:exc}
\begin{center}
\begin{tabular}{cccc}\hline
   & $J^{\pi}$ & $E_X$ (MeV) & \\
\hline
$^7$Li & 1/2$^-$ & 0.4776 & $\Delta E(\pi 1p_{1/2}-\pi 1p_{3/2})$ \\
$^7$Be & 1/2$^-$ & 0.4291 & $\Delta E(\nu 1p_{1/2}-\nu 1p_{3/2})$ \\
$^9$Li & 1/2$^-$ & 2.691 & $\Delta E(\pi 1p_{1/2}-\pi 1p_{3/2})$ \\
$^9$Be & 1/2$^-$ & 2.780 & $\Delta E(\nu 1p_{1/2}-\nu 1p_{3/2})$ \\
\hline
\end{tabular}
\end{center}
\end{table}

\newpage


\begin{table}
\caption{Parameters in the effective potentials for Li nuclei.
The numbers (those in the parentheses) are for neutrons (protons). 
We suppose that the one-neutron separation 
energy of $^{11}$Li is 301 or 350 keV, which is, respectively, denoted by 
$^{11}$Li-a or b in the potential parameters. }
\label{tab:param}
\begin{center}
\begin{tabular}{ccccccc} \hline
       & $U_0$ (MeV) & $r$ (fm) & $a$ (fm) & $U_{LS}$ (MeV) & $a_c$ (fm) \\
\hline
$^6$Li & 51.00 & 1.38 & 0.66 & 1.25 & \\
       & (51.36) & (1.38) & (0.66) & (1.32) & (1.894) \\
$^7$Li & 50.05 & 1.38 & 0.66 & 1.26 & \\
       & (58.10) & (1.38) & (0.66) & (1.32) & (1.902) \\
$^8$Li & 59.30 & 1.00 & 0.53 & 7.27 & \\
       & (88.01) & (1.00) & (0.53) & (9.00) & (1.935) \\
$^9$Li & 55.09 & 1.05 & 0.54 & 12.46 & \\
       & (80.15) & (1.05) & (0.54) & (9.00) & (1.894) \\
$^{11}$Li-a    & 43.81 & 1.08 & 0.51 & 12.46 & \\
                       & (71.72) & (1.08) & (0.51) & (9.00) & (1.894) \\
$^{11}$Li-b    & 40.38 & 1.14 & 0.51 & 12.46 & \\
                       & (66.60) & (1.14) & (0.51) & (9.00) & (1.894) \\
\hline
\end{tabular}
\end{center}
\end{table}
%


\begin{table}
\caption{Comparison of the calculated results to the experimental data
for $^{6-11}$Li nuclei. The numbers in the parentheses 
are the corresponding 
experimental values (see also Tables~\protect\ref{tab:exp} 
and \protect\ref{tab:exc}).} 
\label{tab:fit}
\begin{center}
\begin{tabular}{ccccccc} \hline
       & $S_n$ (MeV) & $S_p$ (MeV) & $\Delta \varepsilon$($n$) (MeV)
& $\Delta \varepsilon$($p$) (MeV) & $r_A$ (fm) \\
\hline
$^6$Li & 5.665 & 4.589 & 0.427 & 0.442 & 2.33 \\
       & [5.664] & [4.589] & [$-$] & [$-$] & [2.32] \\
$^7$Li & 7.246 & 10.006 & 0.429 & 0.476 & 2.33 \\
       & [7.250] & [9.998] & [0.4291] & [0.4776] & [2.33] \\
$^8$Li & 2.033 & 12.449 & 1.397 & 2.643 & 2.37 \\
       & [2.033] & [12.45] & [$-$] & [$-$] & [2.37] \\
$^9$Li & 4.064 & 13.930 & 2.778 & 2.685 & 2.33 \\
       & [4.064] & [13.93] & [2.780] & [2.691] & [2.32] \\
$^{11}$Li-a
       & 0.301 & 15.305 & 2.550 & 2.643 & 3.10 \\
       & [0.326] & [15.30] & [$-$] & [$-$] & [3.10] \\
$^{11}$Li-b
       & 0.350 & 15.301 & 2.747 & 2.737 & 3.10 \\
       & [0.326] & [15.30] & [$-$] & [$-$] & [3.10] \\
\hline
\end{tabular}
\end{center}
\end{table}

\newpage


\begin{table}
\begin{center}
\caption{Calculated energy levels of single-particle states and 
the rms radius of the nucleon. 
The numbers are for neutrons, while those in the parentheses are for 
protons.  $r_N$ stands for the rms radius calculated by the single-particle 
wave function of the nucleon. }
\label{tab:energy}
\begin{tabular}{ccccccccc} \hline
      &  $1s_{1/2}$  & &  &  $1p_{3/2}$  & &  &  $1p_{1/2}$ & \\
      &  $-\varepsilon$ (MeV)  & $r_N$ (fm)  &
      &  $-\varepsilon$ (MeV)  & $r_N$ (fm)  &
      &  $-\varepsilon$ (MeV)  & $r_N$ (fm)  \\
\hline
$^6$Li  & 21.92 & 1.93 & & 5.665 & 2.91 & & 5.237 & 2.95 \\
        & (20.39) & (1.96) & & (4.589) & (2.98) & & (4.147) & (3.02) \\
$^7$Li  & 23.13 & 1.96 & & 7.246 & 2.84 & & 6.817 & 2.87 \\
        & (27.11) & (1.89) & & (10.01) & (2.65) & & (9.530) & (2.67) \\
$^8$Li  & 21.41 & 1.73 & & $2.033$ & 3.15 & & $0.637$ & 4.00 \\
        & (38.51) & (1.51) & & (12.45) & (2.18) & & (9.805) &(2.26) \\
$^9$Li  & $21.94$ & 1.79 & & $4.064$ & 2.84 & & $1.286$ & 3.54 \\
        & (37.49) & (1.59) & & (13.93) & (2.23) & & (11.25) & (2.29) \\
$^{11}$Li-a
 & 17.86 & 1.95 & & 2.851 & 3.13 & & 0.301 & 4.91 \\
 & (36.55) & (1.67) & & (15.30) & (2.27) & & (12.66) & (2.32) \\
$^{11}$Li-b
 & 16.99 & 2.02 & & 3.096 & 3.15 & & 0.350 & 4.82 \\
 & (34.88) & (1.74) & & (15.30) & (2.33) & & (12.56) & (2.38) \\
\hline
\end{tabular}
\end{center}
\end{table}
%


\begin{table}
\caption{Kinetic energy of a nucleon in each shell orbit.
The integration over
the wave function in momentum space was performed in the region
$0 \leq k \leq $ 6 fm$^{-1}$.}
\label{tab:kien}
\begin{center}
\begin{tabular}{cccccccc}\hline
       & $\nu$1$s_{1/2}$ & $\nu$1$p_{3/2}$ & $\nu$1$p_{1/2}$ &
& $\pi$1$s_{1/2}$ & $\pi$1$p_{3/2}$ & $\pi$1$p_{1/2}$  \\
 & (MeV) & (MeV) & (MeV) & & (MeV) & (MeV) & (MeV) \\ \hline
$^6$Li & 12.65 & 16.59 & 16.27 & & 12.31 & 16.08 & 15.72 \\
$^7$Li & 12.26 & 17.01 & 16.77 & & 13.11 & 19.16 & 18.96 \\
$^8$Li & 15.92 & 17.07 & 13.73 & & 20.62 & 28.89 & 27.32 \\
$^9$Li & 14.78 & 18.46 & 14.45 & & 18.59 & 27.23 & 26.03 \\
$^{11}$Li-a
       & 12.50 & 15.66 & 10.73 & & 16.73 & 25.86 & 25.12 \\
$^{11}$Li-b
       & 11.62 & 15.09 & 10.43 & & 15.47 & 24.37 & 23.76 \\
\hline
\end{tabular}
\end{center}
\end{table}

\newpage


\begin{table}
\caption{Calculated average-separation and recoil energies for Li isotopes. 
$\langle \varepsilon \rangle_p$, $\langle \varepsilon \rangle_n$ 
and $\langle \varepsilon \rangle_{tot} 
(= (N \langle \varepsilon \rangle_n 
+ Z \langle \varepsilon \rangle_p)/A)$ are, respectively, for the 
average-separation energies of protons, neutrons and the total system. 
The average-recoil energy is denoted by $\langle T_R \rangle (= 
\langle t \rangle/(A-1))$. 
The bottom row is for the deuteron~\protect\cite{bissey1}. }
\label{tab:average}
\begin{center}
\begin{tabular}{ccccc}\hline
       & $\langle \varepsilon \rangle_p$ (MeV) & 
$\langle \varepsilon \rangle_n$ (MeV) & 
$\langle \varepsilon \rangle_{tot}$ (MeV) & 
$\langle T_R \rangle$ (MeV) \\ \hline 
$^6$Li & 15.1 & 16.5 & 15.8 & 2.3 \\
$^7$Li & 21.4 & 15.2 & 17.9 & 2.1  \\
$^8$Li & 29.8 & 9.78 & 17.3 & 2.4 \\
$^9$Li & 29.6 & 10.0 & 16.6 & 2.1 \\
$^{11}$Li-a
       & 29.5 & 5.97 & 12.4 & 1.4 \\
$^{11}$Li-b
       & 28.4 & 5.88 & 12.0 & 1.3 \\
D & --- & --- & 2.22 & 9.0 \\
\hline
\end{tabular}
\end{center}
\end{table}
%


\newpage


\newpage


\newpage


\newpage


\newpage


%



%



\newpage


\newpage


\newpage


\newpage


\newpage


\newpage


\newpage


\newpage


\end{document}